\documentclass[11pt]{article} 

\usepackage[utf8]{inputenc} 

\usepackage{graphicx} 

\usepackage{amsfonts}

\usepackage{mathtools}
\usepackage{booktabs} 
\usepackage{array} 
\usepackage{verbatim} 
\usepackage{subfig}
\usepackage{setspace}
\usepackage{amsmath}
\usepackage{natbib}
\usepackage{tikz}
\usepackage{accents}

\usetikzlibrary{arrows,decorations.markings}

\usepackage{url}

\usepackage[margin = 1in]{geometry}
\title{A stochastic model for the lifecycle and track of extreme extratropical cyclones in the North Atlantic}

\author{Paul Sharkey$^{1}$, Jonathan A. Tawn$^{2}$ and Simon J. Brown$^{3}$}

\date{} 

\begin{document}
\maketitle
\vspace{-30pt}

\begin{center}
Email: \texttt{pgshky@gmail.com}, \texttt{j.tawn@lancs.ac.uk}, \texttt{simon.brown@metoffice.gov.uk} \\
$^{1}$\emph{JBA Consulting, 1 Belle Vue Square, Skipton, BD23 1FJ, U.K.} \\
$^{2}$\emph{Department of Mathematics and Statistics, Lancaster University, Lancaster, LA1 4YF, U.K.} \\
$^{2}$\emph{Met Office Hadley Centre, Exeter, EX1 3PB, U.K.}
\end{center}

\begin{abstract}
Extratropical cyclones are large-scale weather systems which are often the source of extreme weather events in Northern Europe, often leading to mass infrastructural damage and casualties. Such systems create a local vorticity maxima which tracks across the Atlantic Ocean and from which can be determined a climatology for the region. While there have been considerable advances in developing algorithms for extracting the track and evolution of cyclones from reanalysis datasets, the data record is relatively short. This justifies the need for a statistical model to represent the more extreme characteristics of these weather systems, specifically their intensity and the spatial variability in their tracks. This paper presents a novel simulation-based approach to modelling the lifecycle of extratropical cyclones in terms of both their tracks and vorticity, incorporating various aspects of cyclone evolution and movement. By drawing on methods from extreme value analysis, we can simulate more extreme storms than those observed, representing a useful tool for practitioners concerned with risk assessment with regard to these weather systems.     
\end{abstract}
\hspace{3.5mm}{\bf Keywords:} Extratropical storms, climate extremes, extreme value analysis, serial dependence, spatio-temporal modelling.

\section{Background}
\label{sec:background}
Although the winter climate of western Europe is typically benign, it is often subjected to extreme weather
events characterised by strong winds and heavy rainfall from extratropical cyclones that pose economic, safety and environmental risks. Such events include floods and windstorms that have caused mass infrastructural damage, transport chaos and, in some instances, human fatalities. The storm Desmond, which occurred between 3rd and 8th December 2015, displaced thousands of people from their homes in northern England and Scotland, resulting in an estimated $\pounds$400m worth of damage. \\

Storm Desmond is an example of a synoptic-scale, low-pressure weather system in the North Atlantic Ocean known as an extratropical cyclone. Extratropical cyclones are usually formed as a result of horizontal temperature gradients and develop with a particular lifecycle associated with frontal systems \citep{shapiro1990fronts}. They can be characterised by the paths of local vorticity maxima they generate, which we refer to as tracks. There has been considerable research into cyclone identification and tracking in reanalysis datasets \citep{murray1991numerical, hodges1995feature}. However, this data record is relatively short and thus provides only a limited estimate of the risk from such weather systems, motivating the need for a model to provide improved information on their possible long-term and extreme characteristics. In particular, we would like to know more about the spatial distribution of these storms so that we can identify the regions with more extreme storm activity at a higher level of confidence. We might also like to assess the likelihood of observing more severe storms than those previously observed, where these might occur, and how long these might last. This paper proposes a model from which synthetic storm tracks can be simulated and can be used to perform these assessments in a unified and coherent way. \\

There is limited literature relating to statistical modelling of extratropical cyclones. \citet{sienz2010extreme} used extreme value methods to analyse the effect of climate change on the impact of the North Atlantic Oscillation (NAO) index on cyclone severity. \citet{economou2014spatio} conducted a spatial extreme value analysis of extratropical cyclone pressure minima to estimate probabilities of observing lower-pressure events and the lower endpoint of the distribution of pressure minima. While the model succeeds in capturing the spatial variation of the pressure extremes, it only uses the minimum pressure from a storm track and thus does not account for the spatial and temporal evolution of a cyclone. For example, it may be of interest to practitioners to assess where an extreme storm is likely to propagate. In addition, while the model accounts for the dependence of pressure minima on factors such as its location and the NAO index, it does not explore how it varies relative to the movement of the track. Our approach aims to incorporate both these aspects. \\

Most developments in track modelling have come from the tropical cyclone literature. \citet{casson2000simulation} generated tropical storm tracks by sampling from historical data with random perturbations. Cyclone intensity is modelled dynamically but the history of the process is not incorporated. \citet{rumpf2007stochastic} sampled from kernel density estimates of displacement and direction increments to propagate the track, while \citet{hall2007statistical} use a first-order autoregressive process. Neither incorporate a model for cyclone intensity. Our paper introduces a novel approach to storm track simulation incorporating various properties of extratropical cyclones. This includes the smooth propagation of the track through space, the regional differences between tracks developing at different locations, the tail behaviour of storm intensity and a stochastic termination mechanism.  \\
%
%

Our dataset contains storm track locations at 3-hourly time steps with a vorticity measure associated with each point on the track. Storms are tracked over 36 years (1979-2014) from the ERA-Interim reanalysis dataset described in \citet{dee2011era}. The identification and tracking of the cyclones is performed following the approach used in \citet{hoskins2002new} based on the tracking algorithm described in \citet{hodges1995feature}. Before the identification and tracking progresses the data are smoothed to a resolution of approximately 2.8$^\circ$ and the large-scale background noise is removed. This reduces the inherent noisiness of the vorticity and makes the tracking more reliable for synoptic scale storms. The cyclones are identified by determining the vorticity maxima in the filtered data. Vorticity is preferred to mean surface-level pressure as it has been found to be more suitable for identifying synoptic systems like extratropical storms \citep{hoskins2002new}. Only vorticity values above a threshold of $1.0 \times 10^{-5} s^{-1}$ are considered. Vorticity measurements are linked together through an initial nearest neighbour search that is then refined by constraints on track displacement and smoothness. Storms with a lifespan of less than one day are not considered. Our analysis is focused on storms in an extended winter period (October-March), eliminating any features that may arise due to seasonal effects and focusing on the time of year when storms are considered to be most intense. We restrict our attention to storms passing over the European domain, in particular the region defined in Figure~\ref{fig:e4_region}. We have also removed any Mediterranean storms as these are often influenced by other factors relating to convective behaviour and are not captured well by reanalysis data \citep{akhtar2014medicanes}. \\
%
%

%
%

The paper is structured as follows; Section~\ref{sec:exploratory} details a comprehensive exploratory analysis carried out to assess the main factors influencing storm movement, severity and termination. Section~\ref{sec:methods} describes the methodology to be used in model construction. In Section~\ref{sec:model}, submodels for cyclogenesis, propogation and cyclolysis are outlined, motivated by findings from the exploratory analysis. Section~\ref{sec:discussion} describes the main results based on simulations from the model, followed by some conclusions and opportunities for further work.

\section{Exploratory data analysis}
\label{sec:exploratory}

\begin{figure}
\centering
\includegraphics[width=10cm]{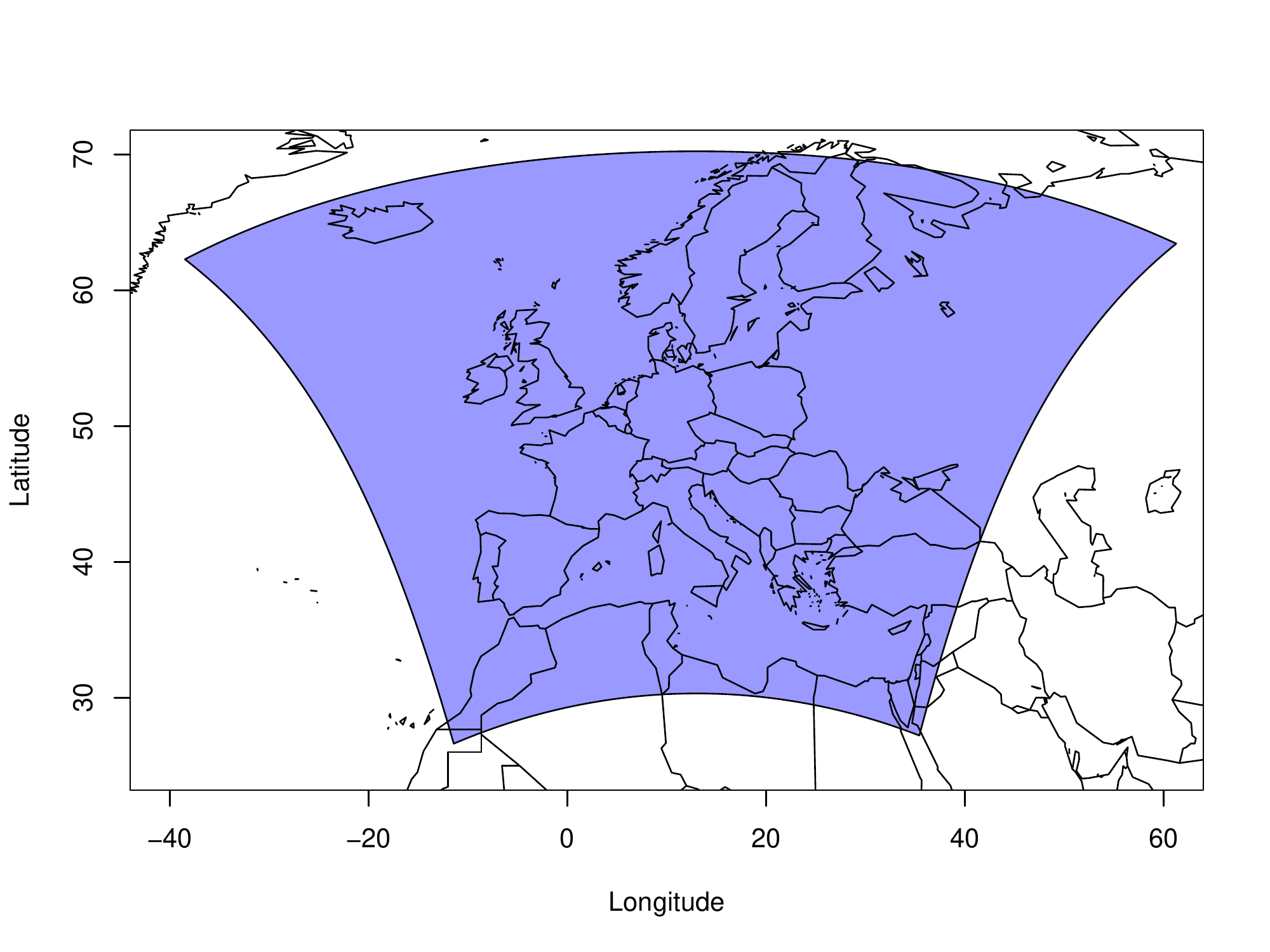}
\caption{The region in which the catalogue of storm tracks cross at some point in their lifetime.}
\label{fig:e4_region}
\end{figure}

\begin{figure}
\centering
\includegraphics[width=10cm]{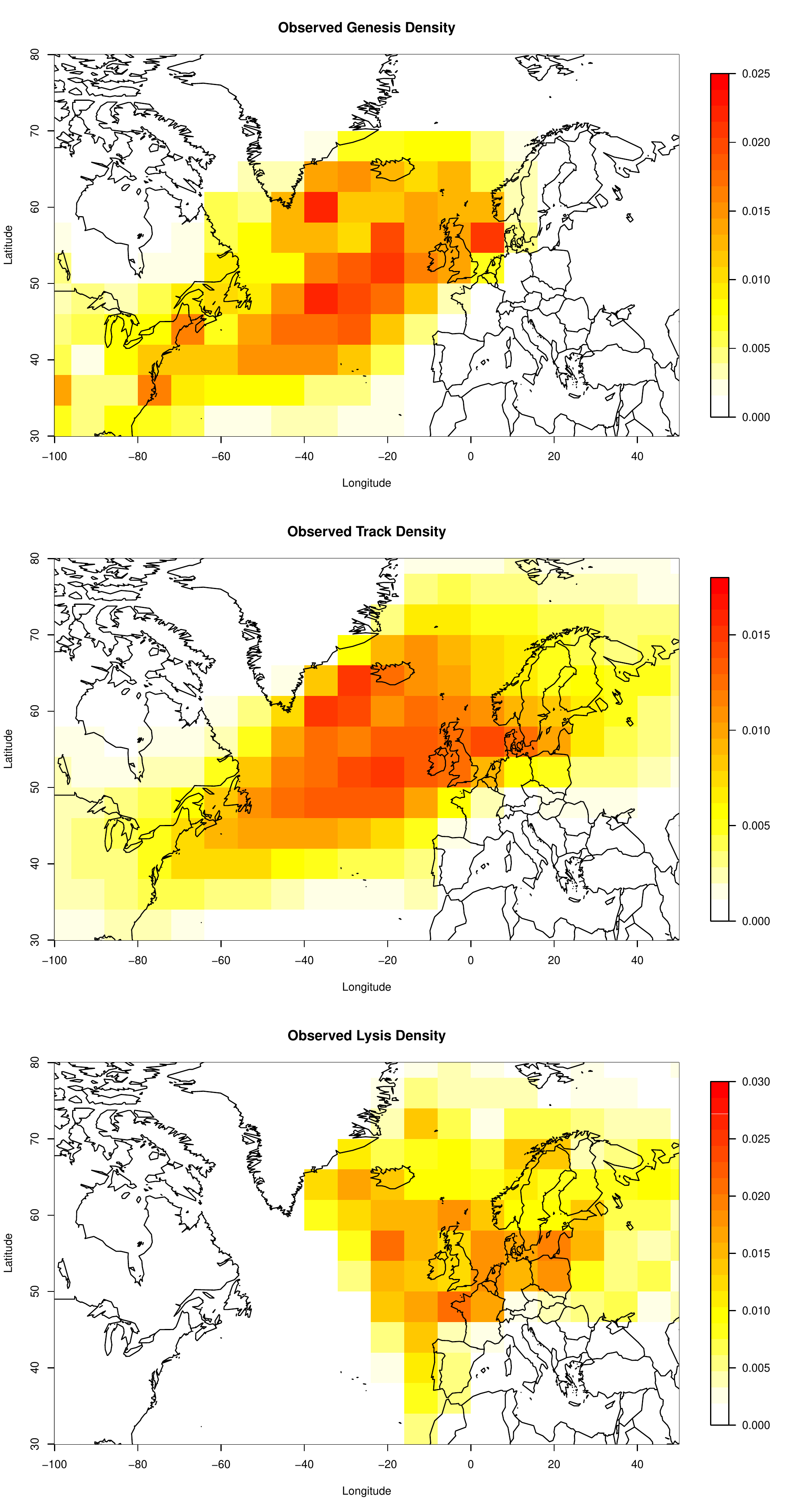}

\caption{Spatial densities of genesis, lysis and overall storm track locations in the analysed dataset.}
\label{fig:spat_dens}
\end{figure}

An extensive exploratory data analysis was carried out in order to gain some intuition regarding the behaviour of storm tracks in the North Atlantic. As discussed in Section 1, our catalogue of observed storm tracks contains only those that have crossed the region shown in Figure~\ref{fig:e4_region}. Our observed set contains $2,944$ storms for the 36 years of data, with approximately 31 observations per storm on average. As these observations are measured at discrete 3-hourly time points, this amounts to the average storm lasting just under four days. Previous analysis of this data \citep{bengtsson2006storm} has shown that these storms tend to begin their existence, known as cyclogenesis, in a corridor across the North Atlantic from south-west to north-east (see Figure~\ref{fig:spat_dens}). Cyclolysis regions, where these storms terminate, tend to be located more towards the eastern Atlantic and Europe. Figure~\ref{fig:spat_dens} also shows the spatial density of all storm track locations; it identifies distinct regions of storm activity in the mid-Atlantic and in the region between Greenland and Iceland. \\ 
%
%


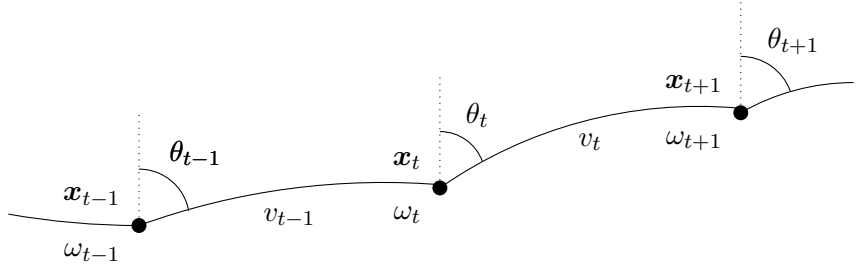
\begin{figure}
\centering
\begin{tikzpicture}
\draw (-4,-0.5)node[circle,fill,inner sep=2pt,label={[label distance=0.05cm]135:$\boldsymbol{x}_{t-1}$},label={[label distance=0.05cm]225:$\omega_{t-1}$}](a){}; 
\draw (0,0)node[circle,fill,inner sep=2pt,label={[label distance=0.05cm]135:$\boldsymbol{x}_{t}$},label={[label distance=0.05cm]225:$\omega_{t}$}](r){};
\draw (4,1) node[circle,fill,inner sep=2pt,label={[label distance=0.05cm]135:$\boldsymbol{x}_{t+1}$},label={[label distance=0.05cm]225:$\omega_{t+1}$}](b){};

\draw[dotted] (a) -- (-4,1);
\draw[dotted] (r) -- (0,1.5);
\draw[dotted] (b) -- (4,2.5);
\draw (a) arc (110:85:9.5);
\draw (r) arc (125:85:6);
\draw (b) arc (120:90:3);
\draw (a) arc (-90:-100:10);

\draw (-4,0.25) arc (90:10:0.67);
\draw (0,0.75) arc (90:20:0.60);
\draw (-2,0) node[label=below:$v_{t-1}$] (){};
\draw (2,1) node[label=below:$v_{t}$] (){};
\draw (-3.25,0) node[label=above:$\theta_{t-1}$] (){};
\draw (0.5,1.4) node[label=below:$\theta_{t}$] (){};
\draw (-3.25,0) node[label=above:$\theta_{t-1}$] (){};
\draw (4.7,2.4) node[label=below:$\theta_{t+1}$] (){};
\draw (4,1.75) arc (90:20:0.7);

\end{tikzpicture}
\caption{Conceptual diagram of the variables extracted from the storm track data.}
%
%
\label{fig:storm_diag}
\end{figure}

\begin{figure}[h!]
\centering
\includegraphics[width=13cm]{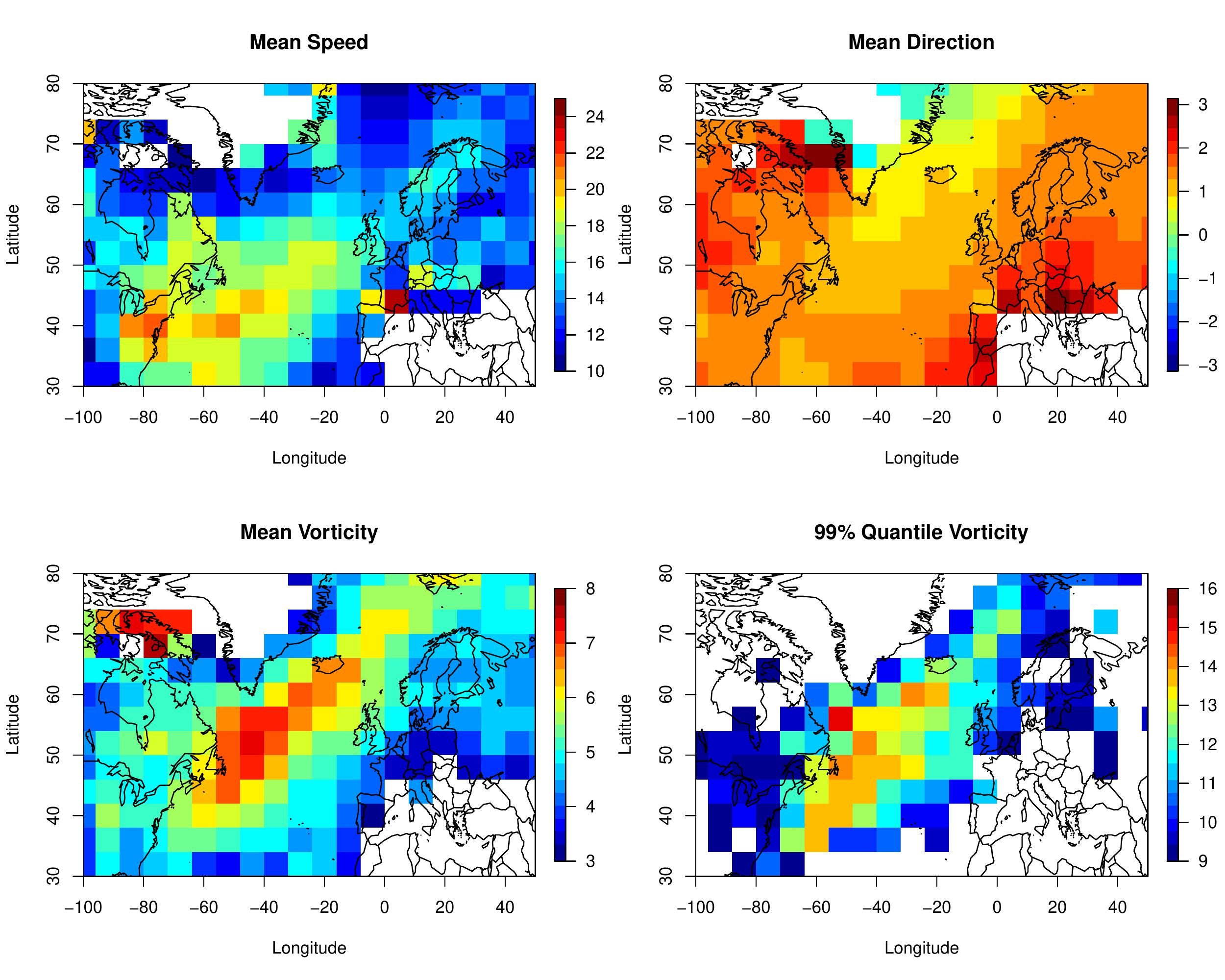}
\caption{Mean speed (top left), mean direction (top right), mean vorticity (bottom left) and 99$\%$ quantile of vorticity (bottom right) in $8^{\circ} \times 4^{\circ}$ grid cells over the spatial domain. For illustration purposes, only 99$\%$ vorticity quantiles of above $9 \times 10^{-5} s^{-1}$ are shown.}
\label{fig:spat_var}
\end{figure}

We extract components of the storm track in order to explore further the variables that determine storm movement and severity. We denote the storm location at time $t$ by $\boldsymbol{X}_t=(X_t,Y_t)$, denoting longitude and latitude coordinates at every 3-hourly interval $t$ respectively. Assuming the Earth is spherical, we derive the speed using distance corresponding to the shortest path between two points along the surface of the sphere, commonly known as the ``great-circle" distance. We denote the speed of the track between $\boldsymbol{X}_{t}$ and $\boldsymbol{X}_{t+1}$ by $V_t$. We choose to model track direction as the initial bearing between $\boldsymbol{X}_{t}$ and $\boldsymbol{X}_{t+1}$, denoted by $\Theta_t \in [-\pi,\pi]$ measured relative to north. We denote the vorticity at location $\boldsymbol{X}_t$ by $\Omega_t$. This variable structure is shown conceptually in Figure~\ref{fig:storm_diag}. \\

These storm variables have a distinct spatial structure (Figure~\ref{fig:spat_var}). The storm tracks tend to begin with an easterly trajectory which becomes more north-easterly as storms move east and to higher latitudes.  Speeds, ranging from $0.07$ to $51.46$ m/s, tend to be highest in an approximate corridor between the eastern coast of the USA and the United Kingdom, with speeds decreasing smoothly as one moves away from this path. The maximum observed vorticity is $15.89 \times 10^{-5} s^{-1}$ with higher vorticities along a similar south-west north-east corridor albeit further north, identified by \citet{bengtsson2006storm} as the highest mean intensity storm regions. Figure~\ref{fig:spat_var} also shows that the strongest of these storms tend to occur in the West Atlantic off the coast of North America.  \\

\begin{figure}[h!]
\centering
\includegraphics[width=12cm]{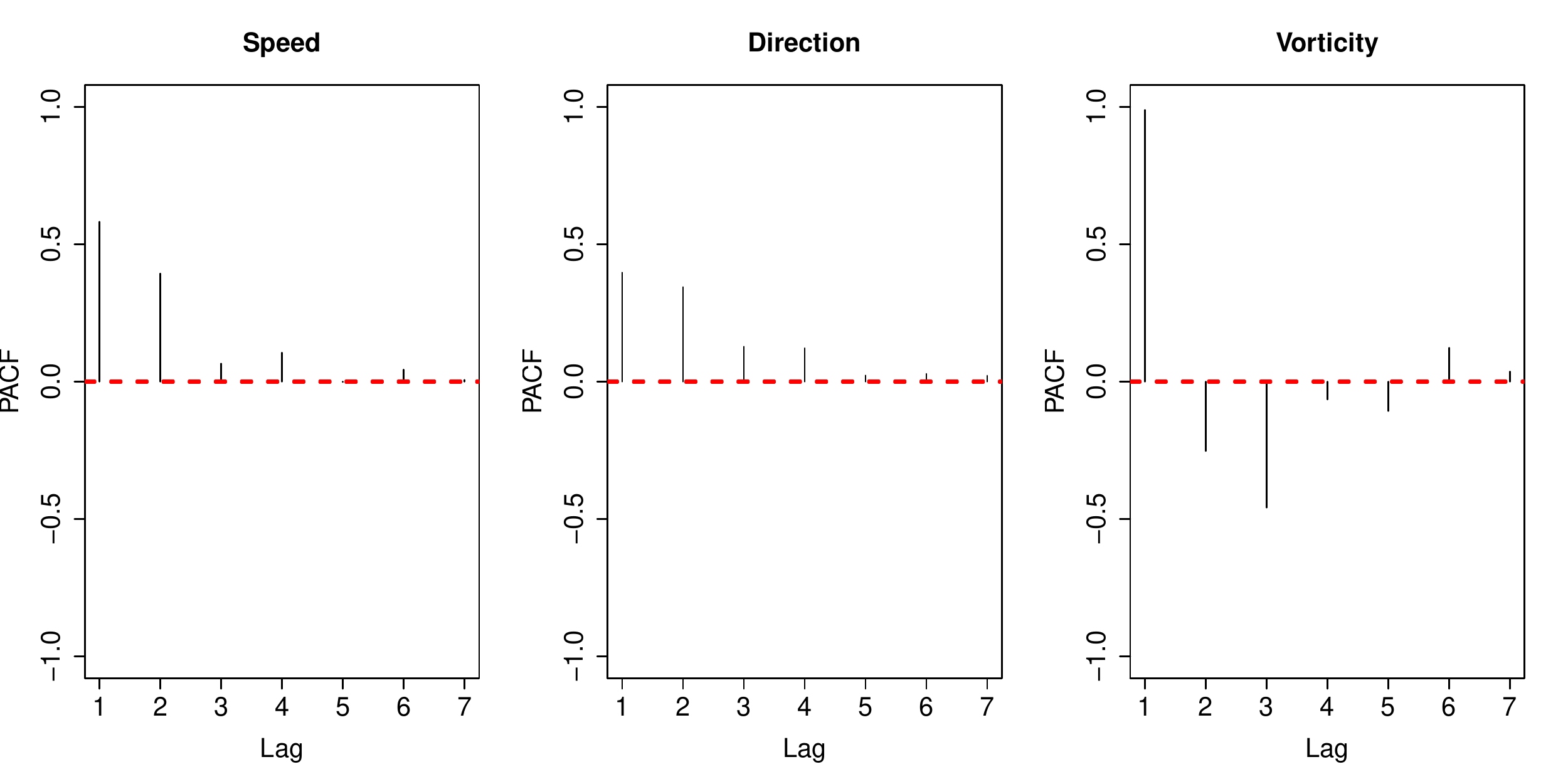}
\caption{Partial autocorrelation plots of speed (left), direction (centre) and vorticity (right).}
\label{fig:pacf_all}
\end{figure}

We investigate the degree of temporal dependence within each variable by examining the partial autocorrelation (PACF) functions for each variable. We identify the order of temporal dependence by the maximum lag at which the PACF is significantly different from $0$. However, because of the size of our dataset, the lowest PACF value that we deem significant is very small, so we interpret the PACF plot by eye to identify a practically relevant order. The PACF plots for $V_t$, $\Theta_t$ and $\Omega_t$ individually are shown in Figure~\ref{fig:pacf_all}; they provide evidence that a third-order relationship for speed, direction and vorticity will capture most of the structure in the data. \\

\begin{figure}[h!]
\centering
\includegraphics[width=12cm]{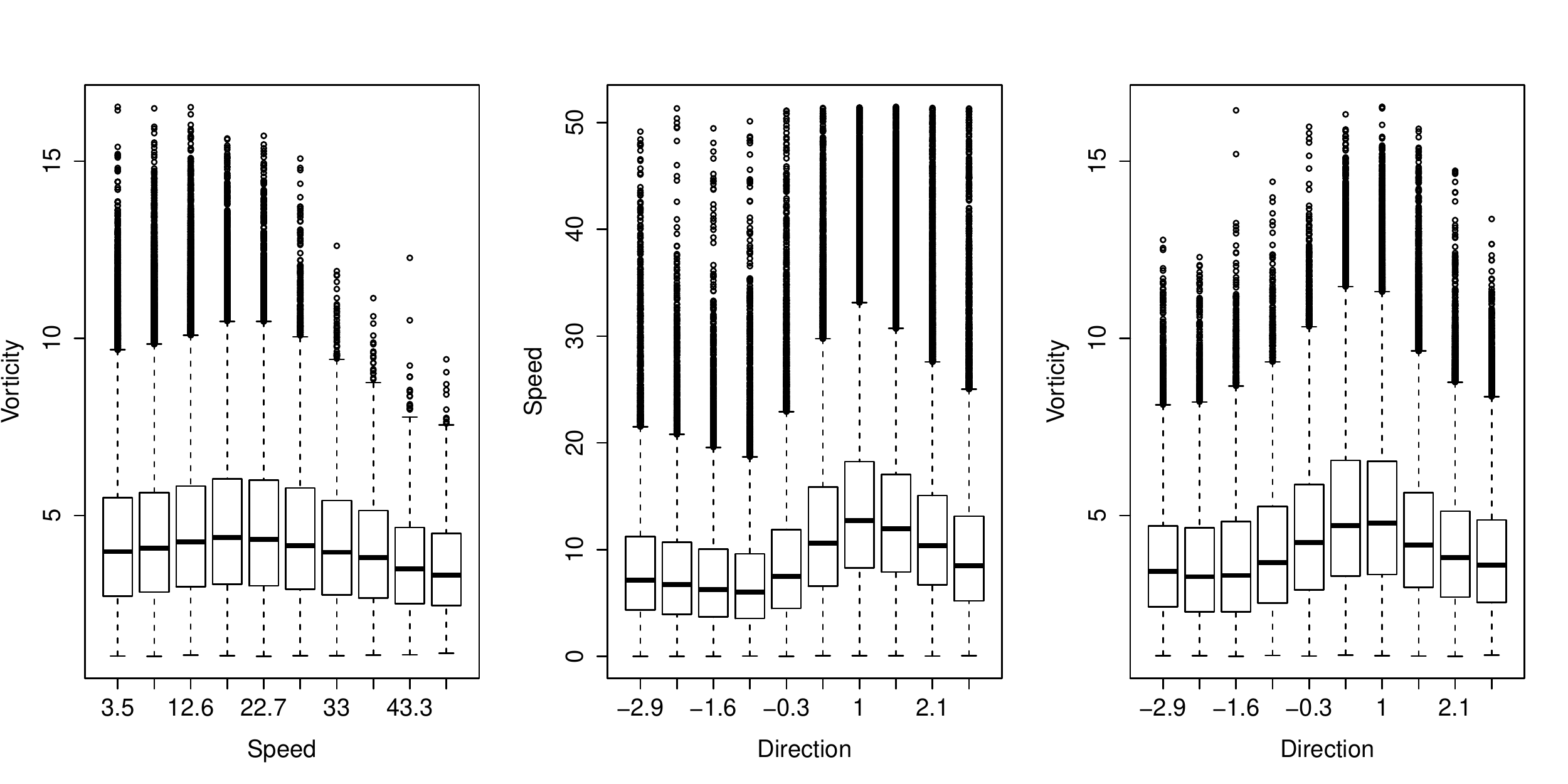}
\caption{The mutual dependence of speed, direction and vorticity shown through boxplots defined by intervals of equal length of the variable on the $x$-axis.}
\label{fig:dependence}
\end{figure}

We explore the possibility that storm intensity and storm movement are interlinked, in other words, that speed, direction and vorticity are dependent. Figure~\ref{fig:dependence} shows that quickly moving and intense storms are more associated with north-easterly/easterly trajectories. It is also clear that a wider range of trajectories are possible when the storm is moving slowly. Figure~\ref{fig:dependence} also suggests that storms will move more slowly when vorticity is large. \\

\begin{figure}[h!]
\centering
\includegraphics[width=10cm]{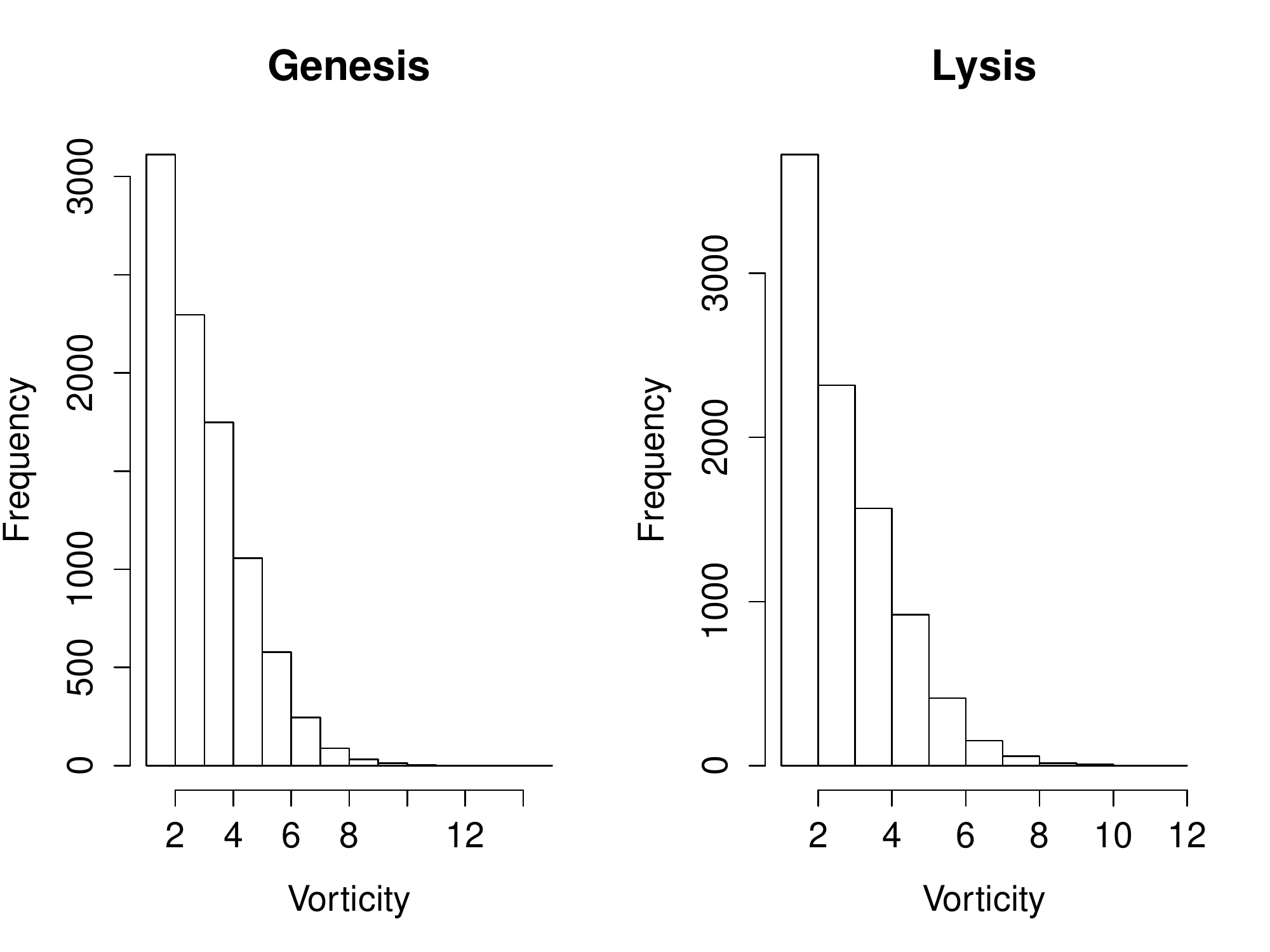}
\caption{The distribution or genesis (left) and lysis (right) vorticities.}
\label{fig:gen_and_lys}
\end{figure}

The feature tracking algorithm of \citet{hoskins2002new} is designed to assign a smooth path to local vorticity maxima above a threshold of $1.0 \times 10^{-5} s^{-1}$. However, genesis and lysis vorticities are generally above this value (see Figure~\ref{fig:gen_and_lys}), which implies either that the storm weakens at a much higher rate or that the tracking algorithm loses the path of the storm. We note that the data may not be representative of the physical termination of a storm, but in the absence of extra information, we design our statistical model to reflect the characteristics of the data. In the context of storm termination, this requires a stochastic mechanism to account for the evident uncertainty in the data. An examination of how the proportion of termination occurrences vary (not shown here) gives evidence to suggest that storms are more likely to terminate if vorticity is low or if the storm is older. Other indicators of storm termination include sharp decreases between consecutive vorticities and the location in space. For example, Figure~\ref{fig:spat_dens} shows that storms are more likely to terminate over western Europe than over Scandinavia. \\  

\section{Methodology}
\label{sec:methods}
\subsection{Introduction}
It is evident from the exploratory data analysis in Section~\ref{sec:exploratory} that extratropical storm tracks are complex systems with many components to be modelled. We use our findings from that analysis to inform a model that represents well the principal physical properties of the storm tracks, in particular, evolution, movement and intensity. We wish to build a model that reflects these processes on a large scale, but also retaining their properties unique to their location in space (see Figure~\ref{fig:spat_var}). We would like the main features of the track to vary smoothly in time. We would also like to extrapolate in order to derive from our model more intense storms than those observed in the data, but to do so requires a rigorous analysis of tail vorticities. We adopt a simulation-based approach to modelling these storm systems, combining sub-models for genesis, propagation and lysis to produce synthetic storm tracks that have the same statistical characteristics as those observed in the data. \\

As our approach aims to propagate the storm in time, it is natural to exploit the time series structure of $\{V_t\}$, $\{\Theta_t\}$ and $\{\Omega_t\}$, which control the movement and severity of a storm. Supported by the exploratory data analysis in Section~\ref{sec:exploratory}, we assume that the multivariate time series $\{(V_t, \Theta_t, \Omega_t)\}$ jointly follows a stationary $k$th order Markov process. By the Markov property, the distribution of the current value of a process is affected only by the previous $k$ time steps of the process. We define an arbitrary $d$-dimensional multivariate time series $\boldsymbol{Z}^{J}_{1:n} = \{ Z_{ij} : i = 1, \hdots, n; j=1, \hdots, d \}$, where $Z_{ij}$ denotes the $i$-th component of the $j$-th dimension and $n$ is the length of the time series. We use the notation $\boldsymbol{Z}^{J}_t$ to denote the tuple at time $t$. We can write the joint density of $\boldsymbol{Z}^{J}_{1:n}$ as
\[ f_{1:n} (\boldsymbol{z}^{J}_{1:n}) = f_{1:k}(\boldsymbol{z}^{J}_{1:k}) \prod^{n-k}_{t=1} f_{k+1|1:k}(\boldsymbol{z}^{J}_{t+k} \mid \boldsymbol{z}^{J}_{t:t+k-1}), \]
where $\boldsymbol{z}^{J}_{p:q} = \{z_{ij} : i = p,\hdots,q ; j=1,\hdots,d \}$, $f_{p:q}$ denotes the joint density function of $\boldsymbol{Z}^{J}_{p:q}$ and $f_{k+1|1:k}(\cdot \mid \cdot)$ is the conditional density function of $\boldsymbol{Z}^{J}_{k+1} \mid \boldsymbol{Z}^{J}_{1:k}$.
This assumption simplifies the modelling process as it becomes only necessary to model the joint distribution of $\boldsymbol{Z}^{J}_{t:t+k}$, which is determined by its marginal distributions and its copula. In some applications, interest lies in estimating probabilities of events beyond the range of the data, for which we draw on methods from extreme value theory.

\subsection{Marginal modelling}
\label{subsec:marginal}
We denote an arbitrary marginal time series component of $\{\boldsymbol{Z}^{J}_t\}$ by $\{Z_t\}$. Under the assumption of stationarity of $\{\boldsymbol{Z}^{J}_t\}$, observations in marginal time series $\{Z_t\}_{t=1}^{n}$ are identically distributed with marginal density function $f$. A simple choice is to model $f$ nonparametrically using the kernel smoothed density function $\hat{f}$, such that
\begin{equation}
 \hat{f}(z) = \frac{1}{nh} \sum_{i=1}^{n} K \left(\frac{z - z_i}{h} \right),
 \label{eq:kernel_density}
 \end{equation}
where $K$ denotes the kernel function, often chosen to be the standard Gaussian density function, and $h$ is the bandwidth.
However, this is known to produce biased estimates in the tails. Instead, for marginal features where the upper tail extremal behaviour is of interest, such as vorticity, we specify a mixture model where the model for the upper tail is motivated through the framework of extreme value analysis. For the remainder of this paper, we denote the quantity $\boldsymbol{Z}_{i:j}= \{Z_l : l = i,\hdots, j \}$. \\

Extreme value analysis is often used to model rare occurrences with the aim of estimating probabilities of events beyond the range of available data. Asymptotic limit models are used in practice as finite-sample approximations for estimating the extreme behaviour of a process. The most widely-used approach is to consider excesses above a suitably high threshold. Under weak conditions on $Z_t$, the distribution of scaled
excesses of a threshold by $Z_t$ converges to the generalised Pareto distribution (GPD) \citep{pickands1975statistical,davison1990models} as the threshold tends to the
upper endpoint $z_F$. This model assumes that the limiting result holds exactly for a large enough threshold $u$. The GPD takes the form
\begin{equation}
\Pr(Z_t-u > z | Z_t > u) = {\left(1+\frac{\xi z}{\psi_{u} }\right)}^{-1/\xi}_{+}, \mbox{                                } z > 0
\label{eq:gpd}
\end{equation}
where $c_{+} = \max(c,0)$ and where $\psi_u > 0$ and $\xi \in \mathbb{R}$ denote the scale and shape parameters respectively. The scale parameter $\psi_u$ is threshold-dependent. A negative shape parameter means that the distribution of excesses has a finite endpoint while values of $\xi = 0$ and $\xi >0$ correspond to exponential- and heavy-tailed distributions respectively. The threshold $u$ is determined using selection diagnostics such as mean residual life plots and checking for threshold stability of $\xi$ \citep{coles2001introduction}. For observations of $Z_t$ larger than a chosen threshold $u$, we replace the kernel estimate defined in~(\ref{eq:kernel_density}) with the GPD model. The marginal model can thus be summarised by
\begin{equation}F(z) = \begin{dcases*} 
							\hat{F} (z) & $z \leq u$ \\
1 - \lambda_u {\left(1+\frac{\xi (z - u)}{\psi_{u}}\right)}^{-1/\xi}_{+} &  $z > u$
										 \end{dcases*}, 
\label{eq:marginal}
\end{equation} \\
where $\hat{F}(z) = \int_{-\infty}^{z} \hat{f}(\gamma) \mathrm{d}\gamma$, where $\hat{f}$ is defined in~(\ref{eq:kernel_density}), and $\lambda_u = 1-\hat{F}(u)$ is the rate of exceedance. A censored maximum likelihood approach is used to obtain estimates of the marginal parameters. For more details on inference for the GPD model, see \citet{coles2001introduction}.

\subsection{Temporal Dependence}
\label{subsec:temp}
Under the Markov assumption, the joint distribution of a time series can be determined by a product of conditional distributions determined by the order of the Markov process. This provides a natural mechanism for propagating a storm in time and incorporating the history of the process. A simple choice of model for $\Pr(Z_{t+k} \leq z \mid \boldsymbol{Z}_{t:t+k-1} = \boldsymbol{z}_{t:t+k-1})$, where $k$ is the order of the Markov process, would be the kernel estimate of the conditional distribution function, the formulation of which can be found in Appendix~\ref{App:sim_kernel}. However, like in the marginal model, this approach poorly captures the temporal dependence structure in the extremes, which is critical when modelling chains of vorticity. This requires an approach for modelling $Z_{t+k} \mid \boldsymbol{Z}_{t:t+k-1}$ in the context of an extreme event, that is, when some functional of $\boldsymbol{Z}_{t:t+k-1}$ exceeds a high threshold $u$. \\

Under the assumption of a stationary $k$th order Markov process, we can model the extremal behaviour of $\{ Z_{t+k} \}$ using the joint distribution of $\boldsymbol{Z}_{t:t+k}$. We can use multivariate extreme value analysis to assess the characteristics of joint tail behaviour with separate models for the marginal and dependence structures. Extremal dependence can be summarised by two broad classes determined by the value of $\chi_{\tau}$, where
\begin{equation}
\chi_{\tau} = \lim_{z \rightarrow z_F} \Pr(Z_{t+\tau} > z | Z_t > z),
\end{equation}
where $\tau \in \mathbb{Z}^{+}$ and $z_F$ is the upper limit of the support of the common marginal distribution. For alternative measures of extremal dependence in a time series context, see \citet{davis2009extremogram} and \citet{ledford2003diagnostics}. A value of $\chi_{\tau} > 0$ refers to the case of asymptotic dependence, where parametric models have been developed with this intrinsic property \citep{coles1999dependence}. Asymptotically independent models, corresponding to the case when $\chi_{\tau} =0$, include contributions by \citet{ledford1996statistics} and \citet{bortot1998models}. Distinguishing between the two classes is crucial as, for example, applying asymptotically dependent models to asymptotically independent data leads to conservative probability estimates of extreme joint events \citep{coles1999dependence}. However, in practice, diagnostics for choosing between the two cases are often highly uncertain. The conditional multivariate extreme value approach of \citet{heffernan2004conditional} is more flexible than standard multivariate models as it covers both cases of asymptotic dependence and asymptotic independence. However, this model gives a limiting representation only for $\boldsymbol{Z}_{t+1:t+k} \mid Z_{t} > u$. To enable a sequential simulation of extremes in time, we draw on methods proposed by \citet{winter2016kth} that model the extremal temporal dependence structure in $\boldsymbol{Z}_{t+1:t+k}$ provided that $Z_t > u$; this approach is described later in this subsection. All vector calculations in this section are to be interpreted componentwise. \\   

After estimation of the marginal model in equation~(\ref{eq:marginal}), it is necessary to transform $\{Z_t\}$ onto common margins to assess extremal dependence after accounting for the marginal structure. Following \citet{keef2013estimation}, we transform onto Laplace margins such that
\begin{equation} S_t = \begin{dcases*}
			\log\{2 F(Z_t)\} & if $Z_t < F^{-1}(0.5)$, \\
			-\log[2\{1-F(Z_t)\}] & if $Z_t \geq F^{-1}(0.5)$,
			\end{dcases*} \label{eq:laplace} \end{equation}
where $\{S_t\}$ denotes the standardised series and $F$ is defined in~(\ref{eq:marginal}).
To explore the conditional distribution $\Pr(\boldsymbol{S}_{t+1:t+m} \leq \boldsymbol{s} \mid S_t > u)$ for large $u$ and integer $m >0$, we use an asymptotically justified form for this distribution as $u \rightarrow \infty$. However, $\boldsymbol{S}_{t+1:t+m}$ requires normalisation so that the limiting conditional distribution is non-degenerate as $u \rightarrow \infty$. \citet{heffernan2004conditional} assume that there exist functions $a:\mathbb{R} \rightarrow \mathbb{R}^{m}$ and $b:\mathbb{R} \rightarrow \mathbb{R}_{+}^{m}$ such that for $s > 0$
\begin{equation}
\Pr\left( \frac{\boldsymbol{S}_{t+1:t+m} - a(S_t)}{b(S_t)} < \boldsymbol{e}_{1:m}, S_t - u > s \, \middle\vert \, S_t > u \right) \rightarrow G_{1:m}(\boldsymbol{e}_{1:m}) \exp(-s),
\label{eq:heff}
\end{equation}
as $u \rightarrow \infty$ with $\boldsymbol{e}_{1:m} \in \mathbb{R}^m$, where $G_{1:m}$ is a joint distribution function that is non-degenerate in each margin. Under weak assumptions on the joint distribution of $\boldsymbol{S}_{t:t+m}$, \citet{heffernan2007limit} show that componentwise $a$ and $b$ must be regularly varying functions
satisfying certain constraints, which for Laplace margins corresponds to each
of the components of $a$ (respectively $b$) being regularly varying functions of index
1 (respectively less than 1). It was found that normalising functions of the simple form 
\[ a(S_t) = \boldsymbol{\alpha}_{1:m} S_t, \hspace{10pt} b(S_t) = {S_t}^{\boldsymbol{\beta}_{1:m}}, \]
where $\boldsymbol{\alpha}_{1:m} \in {[-1,1]}^m$ and $\boldsymbol{\beta}_{1:m} \in {[0,1]}^m$, hold for a very broad range of copulas representing a class of functions which enables parsimonious yet flexible modelling. \citet{winter2016kth} claim that the stationary $k$th order Markov behaviour of $\{S_t\}$ does not impose any constraints on $\boldsymbol{\alpha}_{1:k}$, $\boldsymbol{\beta}_{1:k}$ and $G_{1:k}$, where $k$ is the order of the Markov process, for $k \leq m$. However, $\boldsymbol{\alpha}_{k+1:m}$, $\boldsymbol{\beta}_{k+1:m}$ and $G_{k+1:m}$, for any $m \geq k+1$, are determined entirely by $\boldsymbol{\alpha}_{1:k}$, $\boldsymbol{\beta}_{1:k}$ and $G_{1:k}$ as a result of the stationary Markov behaviour; specific details of this when $k=1$ are described in \citet{papastathopoulos2017extreme}. \\ 

The parameters $\boldsymbol{\alpha}_{1:m}$ and $\boldsymbol{\beta}_{1:m}$ can be used to identify different types of extremal dependence structure. The case of asymptotic dependence between $S_t$ and $S_{t+j}$ corresponds to the case when $\alpha_j = 1$ and $\beta_j = 0$ for $1 \leq j \leq k$, while the case of asymptotic independence arises when $\alpha_j < 1$. Within the asymptotic independence case, positive dependence occurs with $0 < \alpha_j < 1$ or $\alpha_j = 0$ and $\beta_j > 0$; independence when $\alpha_j = \beta_j = 0$ and negative dependence when $-1 \leq \alpha_j < 0$. \\

Our model for the conditional distribution of $\boldsymbol{S}_{t+1:t+k}$ given $S_t > u$ is motivated by the limiting form of the conditional distribution~(\ref{eq:heff}), which we assume is valid for a sufficiently high threshold $u$ and $m=k$. Assuming that $\boldsymbol{S}_{t:t+k}$ has a density, we have that
\begin{equation}
\boldsymbol{S}_{t+1:t+k} | S_t > u = \boldsymbol{\alpha}_{1:k} S_t + {S_t}^{\boldsymbol{\beta}_{1:k}} \boldsymbol{E}_{1:k},
\label{eq:winter}
\end{equation}
where $\boldsymbol{E}_{1:k}$ is a random variable, independent of $t$ and $S_t$, with joint distribution function $G_{1:k}$ and joint density $g_{1:k}$. \citet{winter2016kth} propose an asymptotically motivated heuristic approach to model $S_{t+k}|\boldsymbol{S}_{t:t+k-1}$ when $S_t > u$. Under the assumption that model~(\ref{eq:winter}) holds for $S_t = s_t > u$, it follows that
\begin{equation}
S_{t+k} | (\boldsymbol{S}_{t:t+k-1} = \boldsymbol{s}_{t:t+k-1}) = \alpha_k S_t + {S_t}^{\beta_k} E_{k|1:k-1},
\label{eq:tailchain}
\end{equation}
where $E_{k|1:k-1}$ is a random variable with the same distribution as the conditional distribution of $E_k$ given that
\[\boldsymbol{E}_{1:k-1} = \frac{\boldsymbol{s}_{t:t+k-1} - \boldsymbol{\alpha}_{1:k} s_t}{{s_t}^{\boldsymbol{\beta}_{1:k}}} := \boldsymbol{e}_{1:k-1}. \]
It follows that $S_{t+k+j} | (\boldsymbol{S}_{t:t+k+j-1} = \boldsymbol{s}_{t:t+k+j-1})$, for $j=1,\hdots$ is also given by equation~(\ref{eq:tailchain}), provided $S_{t+j} > u$. We adopt this approach to simulate sequential realisations of an extremal $k$th order Markov process when $S_t > u$. Series generated under this process have negative drifts that ensure the process returns from an extreme state to the body of the distribution, upon which values are generated using the conditional kernel approach outlined in Appendix~\ref{App:sim_kernel}. Dependence parameters $\boldsymbol{\alpha}_{1:k}$ and $\boldsymbol{\beta}_{1:k}$ are estimated using maximum likelihood under the working assumption that $\boldsymbol{E}_{1:k}$ follows a Gaussian distribution. The distribution function $G_{1:k}$ is estimated using the kernel smoothed distribution function of the values of $\boldsymbol{e}_{1:k}$, which are found by inversion of equation~(\ref{eq:tailchain}) under the fitted model. For more details on the inference procedure, see \citet{heffernan2004conditional} and \citet{winter2016kth}.

\section{Simulation Model}
\label{sec:model}

\subsection{Cyclogenesis}
\label{subsec:genesis}
%
%
%
%
%

We construct a model for cyclogenesis conditions using the data observed at the beginning of a storm. In doing so, we would like to model the joint distribution of genesis speed $V_0$, direction $\Theta_0$ and vorticity $\Omega_0$. The spatial variability of $V_t$, $\Theta_t$ and $\Omega_t$ as shown in Figure~\ref{fig:spat_var} is also reflected in the genesis conditions, and thus $(V_0, \Theta_0, \Omega_0)$ should be simulated with respect to genesis location $\boldsymbol{X}_0$. For this reason, we impose an artificial grid on the spatial domain, where each grid cell has dimensions of $8^\circ \times 4^\circ$. This was chosen to be small enough to be able to capture the properties as locally as possible and large enough so that there are enough data to estimate the joint distribution of these properties with sufficient accuracy. We denote $\Delta_{t}$ as the grid cell of the location of the storm track at time $t$. \\

\begin{figure}[h!]
\centering
\includegraphics[width=12cm]{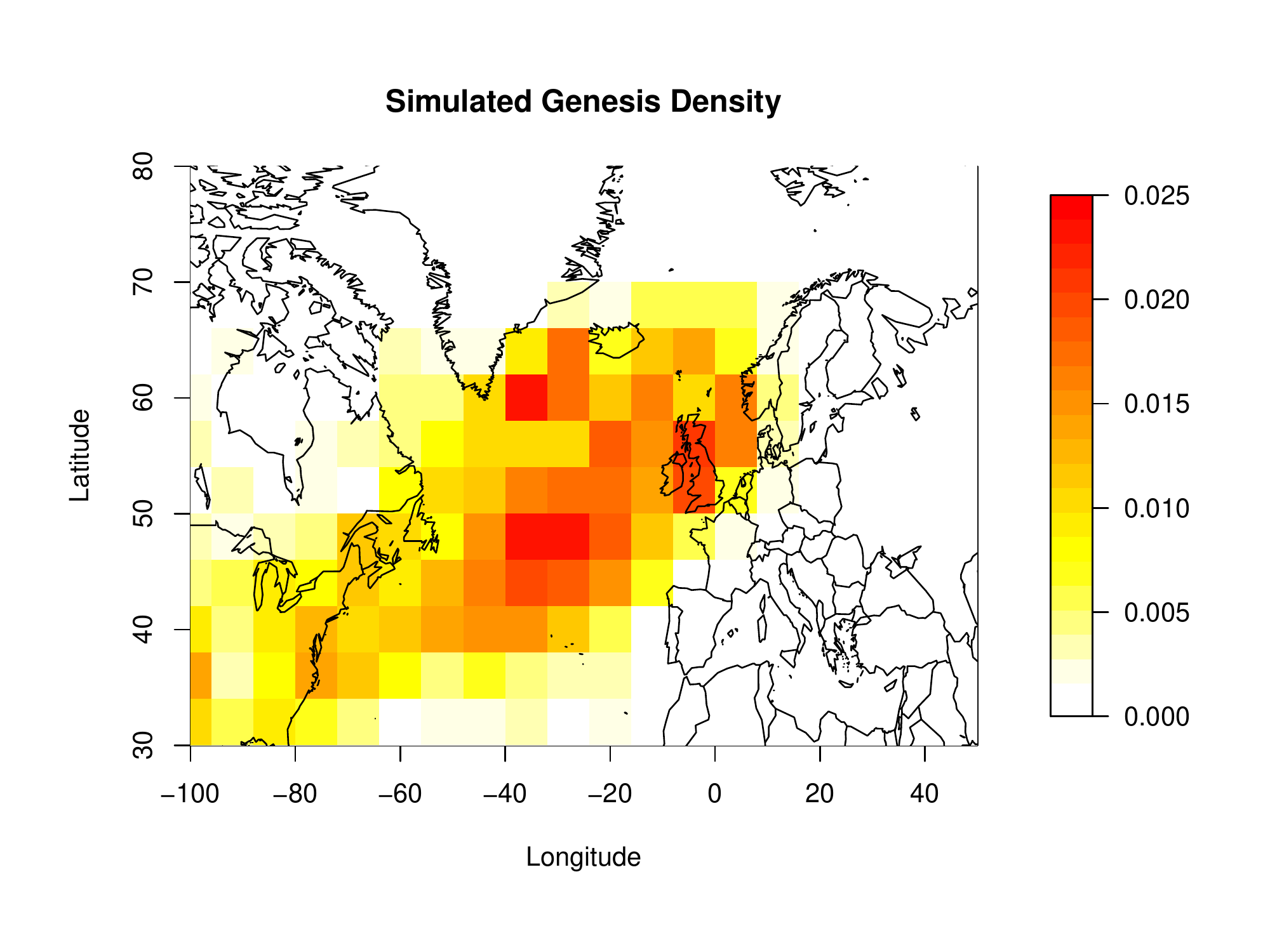}
\caption{Density of the genesis locations of a set of synthetic storm tracks simulated from the model. }
\label{fig:sim_gen}
\end{figure}

We simulate the genesis location from the kernel joint density estimate $\hat{f}(\boldsymbol{x}_0)$, defined in Appendix~\ref{App:sim_kernel}, where the initial locations $\boldsymbol{x}_0$ are the locations from the observed set of storm tracks discussed in Section~\ref{sec:background}. We then use the conditional kernel approach described in Appendix~\ref{App:sim_kernel} to simulate $(v_0, \theta_0, \omega_0)$ jointly from $(V_0, \Theta_0, \Omega_0) \mid \boldsymbol{X}_0 = \boldsymbol{x}_0 \in \Delta_0$.  
We use a Gaussian density kernel function in all cases. We considered a model for $\Theta_0$ using a von Mises kernel to ensure continuity of the density function over $[-\pi,\pi]$. However, we used a non-cyclic Gaussian kernel with repeated shifts of $2\pi$ in the data which produced similar results. We use a correlated kernel for $V_0$, $X_0$ and $\Omega_0$ and an independent for $\Theta_0$, as we believed that a correlated kernel could not sufficiently capture the correlation structure of a cyclic variable. Figure~\ref{fig:sim_gen} shows the density of genesis locations from storm tracks simulated from the model and indicates that our genesis model captures the large scale features quite well whilst having some smaller scale differences such as the southern flank extending a little too far and having a maxima over the UK rather than the North Sea (see Figure~\ref{fig:spat_dens}).
%
%

\subsection{Propagation}

As discussed in Section~\ref{sec:methods}, our exploratory analysis supports the assumption that the storm variables determining movement and severity jointly follow a $k$th order Markov process. As well as temporal dependence, we would like to incorporate any dependence between variables into the propagation scheme in order to represent the joint properties accurately.
The simulated storm track should also reflect the local properties of observed tracks as it moves through space, which the artificial grid introduced in Section~\ref{subsec:genesis} is designed to induce. \\ 
%
%

\begin{figure}[h!]
\begin{center}
\includegraphics[width=16cm]{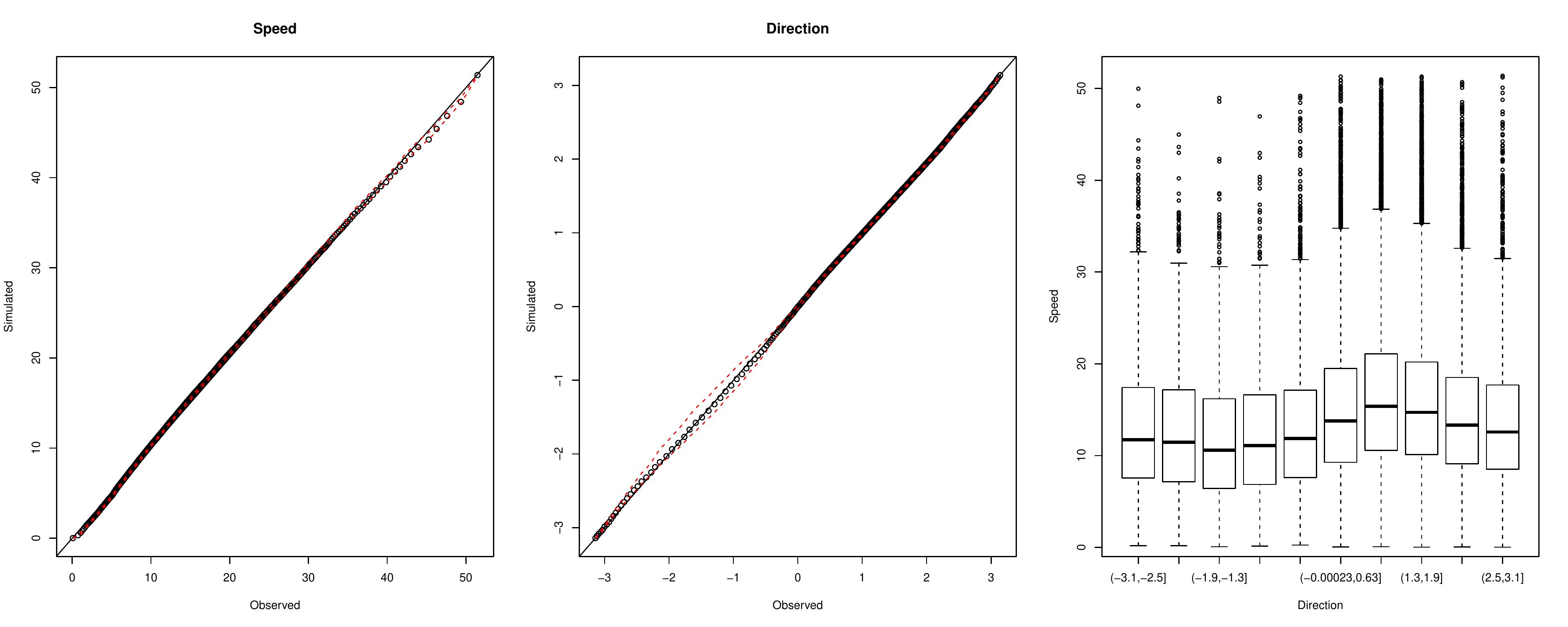}
\end{center}
\caption{QQ plots comparing the observed and simulated marginal distributions of speed (left) and direction (middle) each with $95\%$ tolerance intervals and the dependence between simulated speed and direction shown through a boxplot (right). The simulated data is based on one set of storm tracks drawn from the model of the same number as in the observed set. } 
\label{fig:prop_diag}
\end{figure}

Combining these approaches allows us to construct a joint distribution for $V_t$ and $\Theta_t$, which determine the movement of the storm track, conditional on previous states of the variable, states of other variables, and the grid cell of the storm track location. For all times $1 \leq j \leq k$, where $k$ is the order of the Markov process jointly for $\{(V_t,\Theta_t,\Omega_t)\}$, we simulate:
%
%
%
\begin{eqnarray}
\theta_j &\sim& \Theta_j \mid \Theta_{0:j-1}= \theta_{0:j-1}, \boldsymbol{x}_{j} \in \Delta_{{j}} \nonumber\\
v_j &\sim& V_j \mid V_{0:j-1} = v_{0:j-1}, \Theta_{j} = \theta_{j}, \boldsymbol{x}_{j} \in \Delta_{{j}} \label{eq:cond1}
\end{eqnarray}
When $j>k$, we simulate:
\begin{eqnarray}
\theta_j &\sim& \Theta_j \mid \Theta_{j-k:j-1}=\theta_{j-k:j-1}, \boldsymbol{x}_{j} \in \Delta_{{j}} \nonumber\\
v_j &\sim& V_j \mid V_{j-k:j-1} = v_{j-k:j-1}, \Theta_j = \theta_j, \boldsymbol{x}_{j} \in \Delta_{{j}}.\label{eq:cond2}
\end{eqnarray}
Simulated values $v_j$ and $\theta_j$ are obtained from the kernel estimate of the conditional distribution in (\ref{eq:cond1}) and (\ref{eq:cond2}) as discussed in Section~\ref{subsec:temp} and formulated in Appendix~\ref{App:sim_kernel}. The exploratory analysis in Section~\ref{sec:exploratory} suggests that $k=3$ is an appropriate choice. The dependence between speed and direction is induced through conditioning $V_j$ on $\Theta_j$. As in Section~\ref{subsec:genesis}, we used a Gaussian kernel function in both cases. When simulating $\theta_j$, we again considered a von-Mises kernel to ensure continuity of the density function over $[-\pi,\pi]$, but a non-cyclic Gaussian kernel with repeated shifts of $2\pi$ in the data produced similar results. We use the simulated values to propagate the storm. Longitude and latitude coordinates $\boldsymbol{x}_{j+1} = (x_{j+1}, y_{j+1})$ are calculated using the formula:
\begin{eqnarray}
y_{j+1} &=& \sin^{-1} \left(\sin(y_{j}) \cos\left(\frac{v_j \nabla_j}{R}\right) + \cos(y_{j}) \sin\left(\frac{v_j \nabla_j}{R}\right) \cos(\theta_j) \right); \nonumber \\
x_{j+1} &=&  x_{j} + \text{Tan}^{-1} \left( \sin(\theta_j) \sin\left(\frac{v_j \nabla_j}{R}\right) \cos(y_{j}) \cos\left(\frac{v_j \nabla_j}{R}\right) - \sin(y_{j}) \sin(y_{j+1}) \right), \nonumber
\end{eqnarray}
where $(x_{j},y_{j})$ denote the longitude and latitude coordinates at time $j$, $R$ denotes the radius of the Earth, taken to be 6371 km, $\nabla_j$ denotes the time difference in seconds between time $j$ and $j+1$ and Tan$^{-1}$ denotes the four-quadrant inverse tangent function. If at time $j+1$, a simulated track enters a region such that $\Delta_{j+1}$ has seen no observed storm activity, the track is reverted to time $j$, giving the algorithm 10 opportunities to find a trajectory towards a grid cell that has observed storm activity. If no such trajectory is found, the storm is terminated. The QQ plot in Figure~\ref{fig:prop_diag} shows that the simulation model replicates exceptionally well the observed marginal distributions of speed and direction. Figure~\ref{fig:prop_diag} also shows the model captures the tendency of storm tracks to move more quickly in a northeasterly direction (see Figure~\ref{fig:dependence}).  \\

%
%
%
%

\subsection{Vorticity modelling}
\label{subsec:vorticity}
The relationship between the vorticity of a storm track and its influence on the weather is complex. Data analysis (not shown here) demonstrates that vorticity is weakly correlated with the maximum wind speed observed in the vicinity of the track, as well as showing evidence that large spatial wind speed events are linked with large vorticities. Therefore it is critical to model carefully the spatial and temporal characteristics of this variable as well as its behaviour in the extremes. While the kernel approach is useful at simulating realistic chains of $V_t$ and $\Theta_t$, the marginal distribution and dependence structure is estimated using the entire series and may therefore lead to bias in the extremes. In practice, we would like to estimate probabilities of observing storms with a vorticity not yet observed. We use asymptotically justified limit models from extreme value theory to estimate these probabilities using the observed extreme events. Our simulation method combines the kernel approach used in~(\ref{eq:cond1}) and~(\ref{eq:cond2}) with techniques outlined in Sections~\ref{subsec:marginal} and~\ref{subsec:temp} for tail models with extremal temporal dependence structure.     \\

The exploratory analysis in Section~\ref{sec:exploratory} shows how the upper tail of vorticity varies with respect to the movement and location of the track. To account for this in an extreme value model, we first use the preprocessing method of \citet{eastoe2009modelling} to transform the data to approximate stationarity. Specifically, we use a Box-Cox-location-scale model of the form
\begin{equation}
\frac{\Omega_t^{\lambda} -1}{\lambda} = \mu(\boldsymbol{\nu_t}) + \sigma(\boldsymbol{\nu_t}) W_t,
\label{eq:preprocess}
\end{equation}
where $W_t$ is assumed to be approximately stationary, $\lambda$ denotes the Box-Cox parameter and $\sigma$ and $\mu$ are functions of a vector of covariates $\boldsymbol{\nu}_t$. For the purpose of inference on $\lambda$, $\mu$ and $\sigma$, $W_t$ is assumed to be $\mathcal{N}(0,1)$. Parameter estimates are obtained using maximum likelihood. We fit this model to data within the longitude range $(-60^{\circ},20^{\circ})$ and latitude range $(40^{\circ},80^{\circ})$ as our interest lies in the extremal behaviour of vorticity in this region. A number of combinations of covariates were considered and the model fit was assessed using likelihood ratio testing. The best-fitting model features functions of latitude, longitude, direction and speed in both $\mu$ and $\sigma$. This ensures that the variation in large vorticities over space is captured (see Figure~\ref{fig:spat_var}) while also ensuring the dependence structure shown in Figure~\ref{fig:dependence} holds for large vorticity values.\\

We model the excesses of $W_t$ above some suitably high threshold $u$ using the GPD tail model as discussed in Section~\ref{subsec:marginal}. A threshold of $u=1.5$ is selected, corresponding to the $98.13\%$ quantile of $W_t$. The maximum likelihood estimates are $\hat{\psi}_u = 0.449$ and $\hat{\xi} = -0.246$. Note that the negative shape parameter estimate implies a physical upper limit to the vorticity distribution. This is consistent with the extremal analysis of mean surface level pressure in \citet{economou2014spatio}, as vorticity and mean surface level pressure tend to behave similarly in the context of extratropical storms \citep{hoskins2002new}. Desired tail quantiles of $\Omega_t$ are determined by back-transformation. \\
%
%
 
The temporal propagation of vorticity is set out as follows. We describe separately the cases for simulating realisations of $\Omega_j$, for some arbitrary time $j$, given that the $k$ previous observations of $W_j$ (a function of $\Omega_j$) are in non-extreme and extreme states respectively.  First, consider the process when $\Omega_j$ is such that the $k$ previous observations of $W_j$ are in a non-extreme state. In particular, we consider two cases. For all times $1 \leq j \leq k=3$ and $\boldsymbol{\Omega}_{0:j-1} = \boldsymbol{\omega}_{0:j-1}$ such that $\max\{\boldsymbol{W}_{0:j-1}\} < u$, we simulate
\[\omega_j \sim \Omega_j \mid \boldsymbol{\Omega}_{0:j-1}=\boldsymbol{\omega}_{0:j-1}, \Theta_{j-1}=\theta_{j-1}, \boldsymbol{x}_{j} \in \Delta_{j}.\]
Next, consider when $j > k$ and $\boldsymbol{\Omega}_{j-k:j-1}=\boldsymbol{\omega}_{j-k:j-1}$ such that $\max\{\boldsymbol{W}_{j-k:j-1}\} < u$. In this case, we simulate
\[\omega_j \sim \Omega_j \mid \boldsymbol{\Omega}_{{j-k}:{j-1}}=\boldsymbol{\omega}_{{j-k}:{j-1}}, \Theta_{j-1} = \theta_{j-1}, \boldsymbol{x}_{j} \in \Delta_{j}. \]
Empirical evidence suggests that vorticity and track speed are approximately independent conditional on the bearing, and since speed is simulated with this conditioning in~(\ref{eq:cond1}) and~(\ref{eq:cond2}), we believe that simulating $\omega_j$ conditional on $\theta_{j-1}$ is sufficient to represent the dependence between storm movement and intensity. The conditional distribution $\Omega_j \mid \cdot$ is estimated using the kernel approach described in Appendix~\ref{App:sim_kernel}.  \\
%
%
%

\begin{figure}[h!]
\centering
\includegraphics[width=9cm]{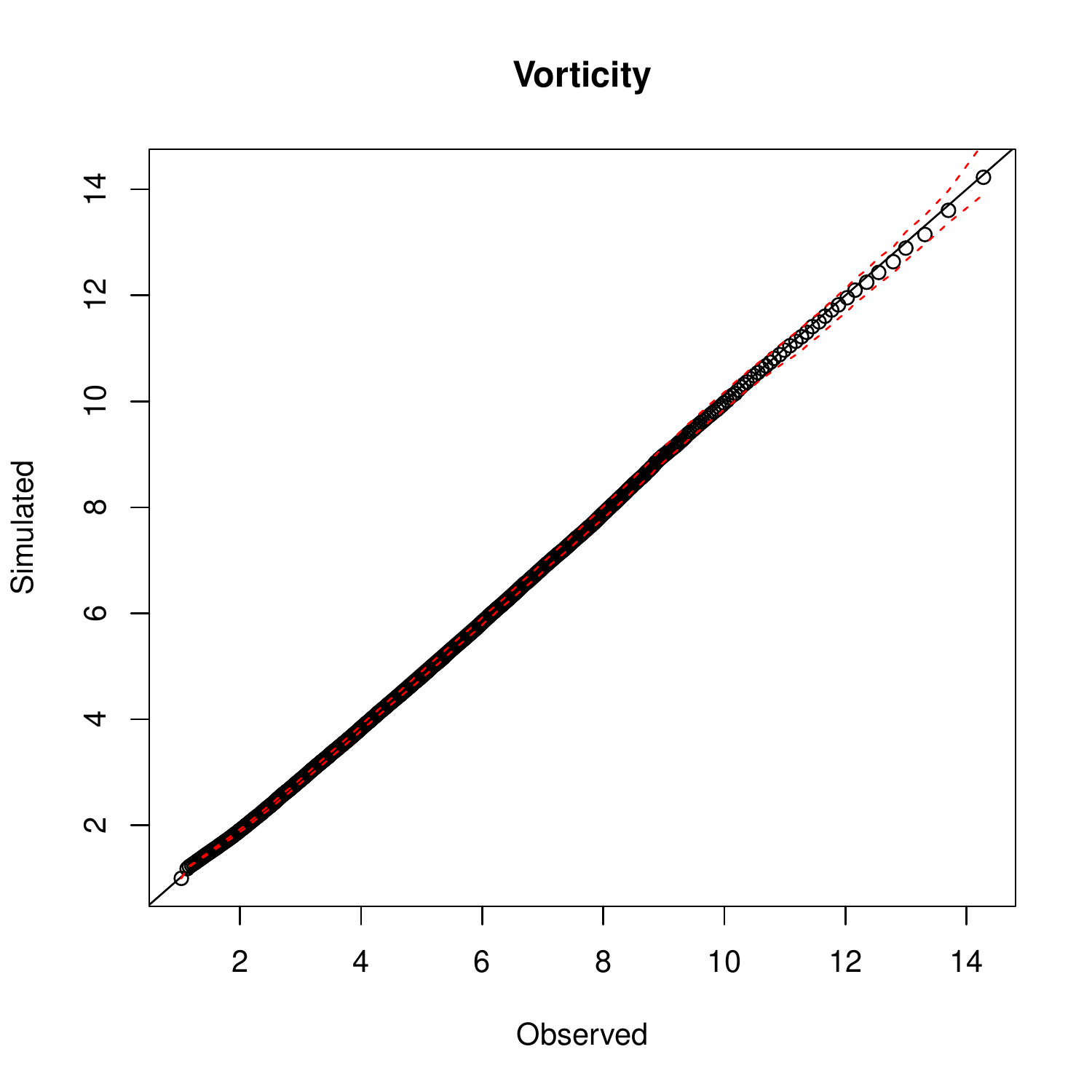}
\caption{QQ plot (with 95$\%$ tolerance intervals) comparing observed marginal distributions to vorticities from a set of simulated storm tracks of the same number as those in the observed set.}
\label{fig:vor_qq}
\end{figure}

Next, when $\Omega_j$ is such that the previous $k$ observations of $W_j$ are in an extreme state, we adopt the model of \citet{winter2016kth} for simulating tail chains under the assumption of an extremal $k$th order Markov process. In particular, we transform the preprocessed series $W_t$ onto Laplace margins as in~(\ref{eq:laplace}), denoting the transformed quantity by $S_t$. Provided at least one of the last $k$ observations previous to $W_j$ is in an extreme state, we simulate realisations of $S_j$ using the tail chain approach before backtransforming to obtain a realisation of $\Omega_j$. To be precise, let $l$ be the number of consecutive excesses of $\{W_t\}$ above $u$ previous to time $j$, such that
\[ l_j = \max\{ i \in \{1,\hdots,k\} : \min\{\boldsymbol{W}_{j-i:j-1}\} > u \}. \]
For example, if $S_{j-3}$ and $S_{j-2}$ are less than $u$ but $S_{j-1} > u$, we use a first order structure to simulate $S_j$. Therefore $l_j$ represents the order to be used in the simulation of $S_j$.
After determining the order $l=l_j$, we then simulate:
\[S_j = \hat{\alpha}_{l} S_{j-l} + {(S_{j-l})}^{\hat{\beta}_{l}} e_{j \mid {j-l+1}:{j-1}}, \] 
where $(\hat{\alpha}_{l},\hat{\beta}_{l})$ denote the maximum likelihood estimates of the dependence parameters and $e_{j \mid {j-l+1}:{j-1}}$ is sampled independently from $\hat{G}_{j \mid {j-l+1}:{j-1}}$. The value $S_j$ is transformed to obtain the preprocessed $W_j$ by inverting equation~(\ref{eq:laplace}). Vorticity $\Omega_j = \omega_j$ is then obtained by inverting equation~(\ref{eq:preprocess}). The QQ plot in Figure~\ref{fig:vor_qq} shows that the model captures well both the body of the vorticity distribution and its extremes.

\subsection{Cyclolysis}
\label{subsec:lysis}
The termination of a storm and its track, termed cyclolysis, is not well defined in the observed track dataset, as discussed in Section~\ref{sec:exploratory}. There are a number of instances where a storm terminates at a value of vorticity that is significantly larger than the critical threshold defined by the 
tracking algorithm, suggesting storms can fade quickly. This motivates the need for a stochastic termination mechanism to be applied to the simulated storm, as discussed in Section~\ref{sec:background}. The exploratory analysis in Section~\ref{sec:exploratory} suggests several factors influence the risk of termination, including vorticity, age and location. We model these covariate effects using a logistic generalised additive model \citep{wood2006generalized}. We estimate a probability of termination at each simulated 3-hourly time step of the storm for $t \geq 8$ in order to replicate the constraint of the tracking algorithm only to consider storms that last for at least 24 hours. The termination mechanism is implemented after the storm track enters the region shown in Figure~\ref{fig:e4_region}. Let $T_{t}$ be a Bernoulli random variable such that:
\[
 T_{t} = \begin{dcases*}
        1  & when the storm terminates at time $t$\\
        0 & otherwise
        \end{dcases*}
\]
So $T_{t} \sim$ Bernoulli$(p_t)$, where
\[ p_t = \begin{dcases*}
		  0 & $t < 8$ \\
		  \frac{\exp\left\{ \sum_{i=1}^{q} s_i (\nu_{i,t}) \right\}}{1+\exp\left\{\sum_{i=1}^{q} s_i (\nu_{i,t}) \right\}} & $t \geq 8$
		  \end{dcases*}
\]
where $s_i$ is a smooth non-linear function of covariate $\nu_i$ with $i \in (1,\hdots,q)$, where $q$ is the number of covariates. The smooth functions are represented by penalised regression splines, where the smoothing parameter is determined using generalised cross validation (GCV) and the model is fitted using a penalised maximum likelihood formulation. For more details on additive models, see \citet{wood2006generalized}. \\

\begin{figure}[h!]
\begin{center}
\includegraphics[width=17cm]{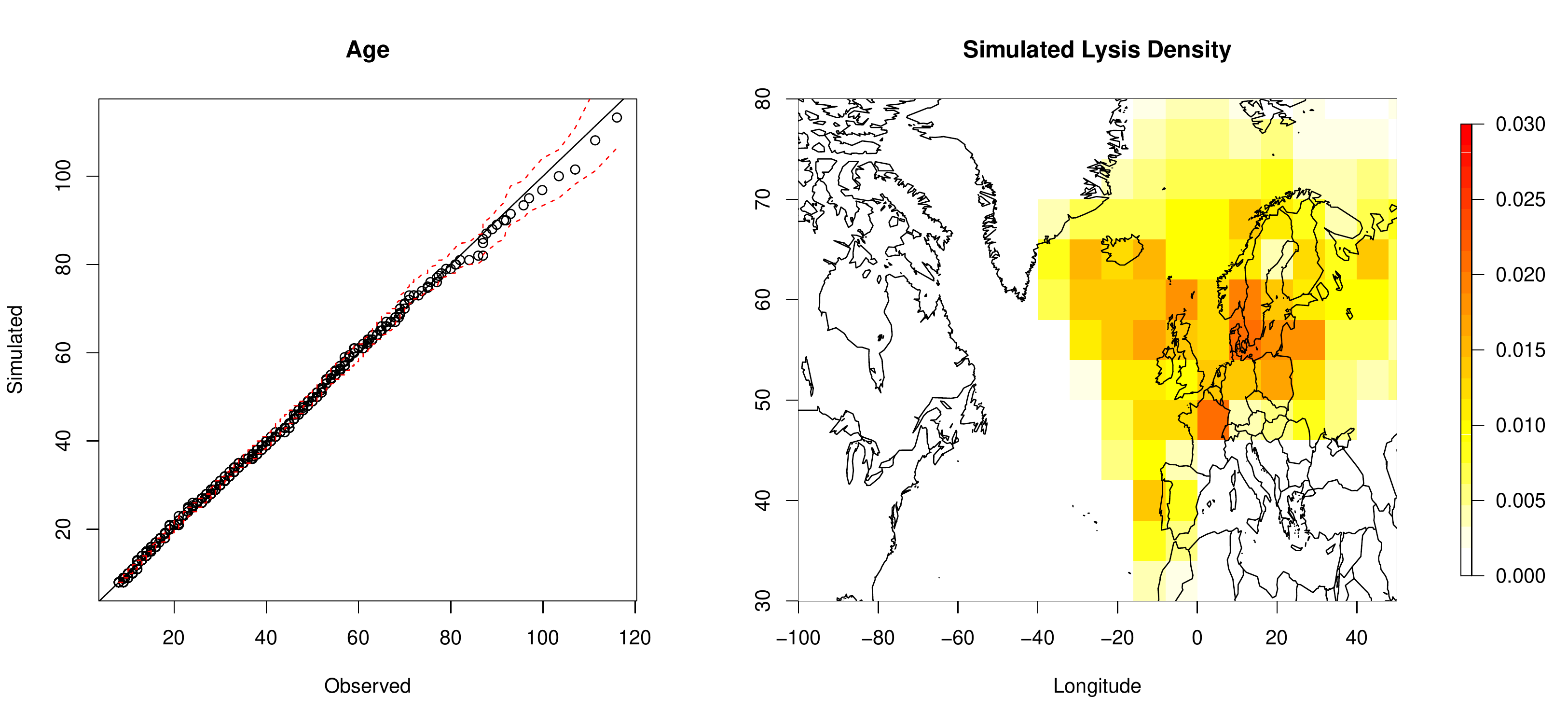}
\end{center}
\caption{QQ plot (with $95\%$ tolerance intervals) comparing observed storm lifetimes with lifetimes of storms simulated from the model (left). One unit of age is defined as one 3-hourly interval. Spatial density of storm lysis locations (right) based on a set of storm tracks simulated from the model of the same number at those in the observed set.} 
\label{fig:age_qq}
\end{figure}

The effect of the covariates on the model fit was assessed using AIC. The best-fitting model under this criterion consisted of functions of several variables including vorticity, age, longitude and latitude. As hypothesised based on the exploratory data analysis, the fitted model shows that a track is more likely to terminate if the vorticity is low or the storm has experienced a large sudden reduction in vorticity. A track is also more likely to terminate if it is older. The QQ plot in Figure~\ref{fig:age_qq} shows that storm lifetimes are being well captured by the model, while the spatial density of lysis locations compares well with the observed (see Figure~\ref{fig:spat_dens}). In both cases, storms tend to terminate over the northeast Atlantic and northwest Europe.

\subsection{Risk analysis}
\label{subsec:risk}
As discussed in Section~\ref{subsec:vorticity}, the relationship between the extreme weather impact caused by extratropical storms is complex and warrants further investigation. As the vorticity of a storm is known to be correlated with characteristics of large wind speed events, it is useful for practitioners to be aware of the rate and size of extreme vorticity events in different regions. We can estimate the probability of such events through Monte Carlo simulation. For illustration purposes, we fix our region of interest to be a longitude and latitude range containing the UK; in particular, we define the region $\Gamma =  \{(x,y): x \in(-11^\circ,2^\circ); y \in (50^\circ,60^\circ) \}$. To estimate the probability of exceeding a vorticity $\omega$ in this region, we calculate
\begin{equation} \hat{\Pr}(\Omega_t > \omega\mid \boldsymbol{X}_t \in \Gamma) = \frac{\sum_{i=1}^{N} \sum_{j=1}^{n_i} \mathbb{I} \{ \omega_{ij} > \omega, \boldsymbol{x}_{ij} \in \Gamma \}}{ \sum_{i=1}^{N} \sum_{j=1}^{n_i} \mathbb{I} \{\boldsymbol{x}_{ij} \in \Gamma \} } ,
\label{eq:risk_prob}
\end{equation}
where $\mathbb{I}$ is the indicator function and $\boldsymbol{x}_{ij}$ and $\omega_{ij}$ denote the location and vorticity respectively at the $j$th time step of the $i$th storm, $n_i$ denotes the time length of storm $i$ and $N$ denotes the number of simulated storms. One could alternatively characterise a risk measure in terms of the maximum vorticity of a storm, denoted $\omega_{\text{max}}$, such that
\[ \hat{\Pr}(\Omega_t > \omega_{\text{max}} \mid \boldsymbol{X}_t \in \Gamma) = \frac{\sum_{i=1}^{N}  \mathbb{I} \{ \max_j \omega_{ij} > \omega_{\text{max}}, \boldsymbol{x}_{ij} \in \Gamma \}}{ \sum_{i=1}^{N}  \mathbb{I} \{ \max_j \boldsymbol{x}_{ij} \in \Gamma \} } ,\]
which would remove the possibility that multiple excesses could be observed in the same storm. For the purpose of illustration, however, we continue with the characterisation in~(\ref{eq:risk_prob}).\\

\begin{figure}
\centering
\includegraphics[width=12cm]{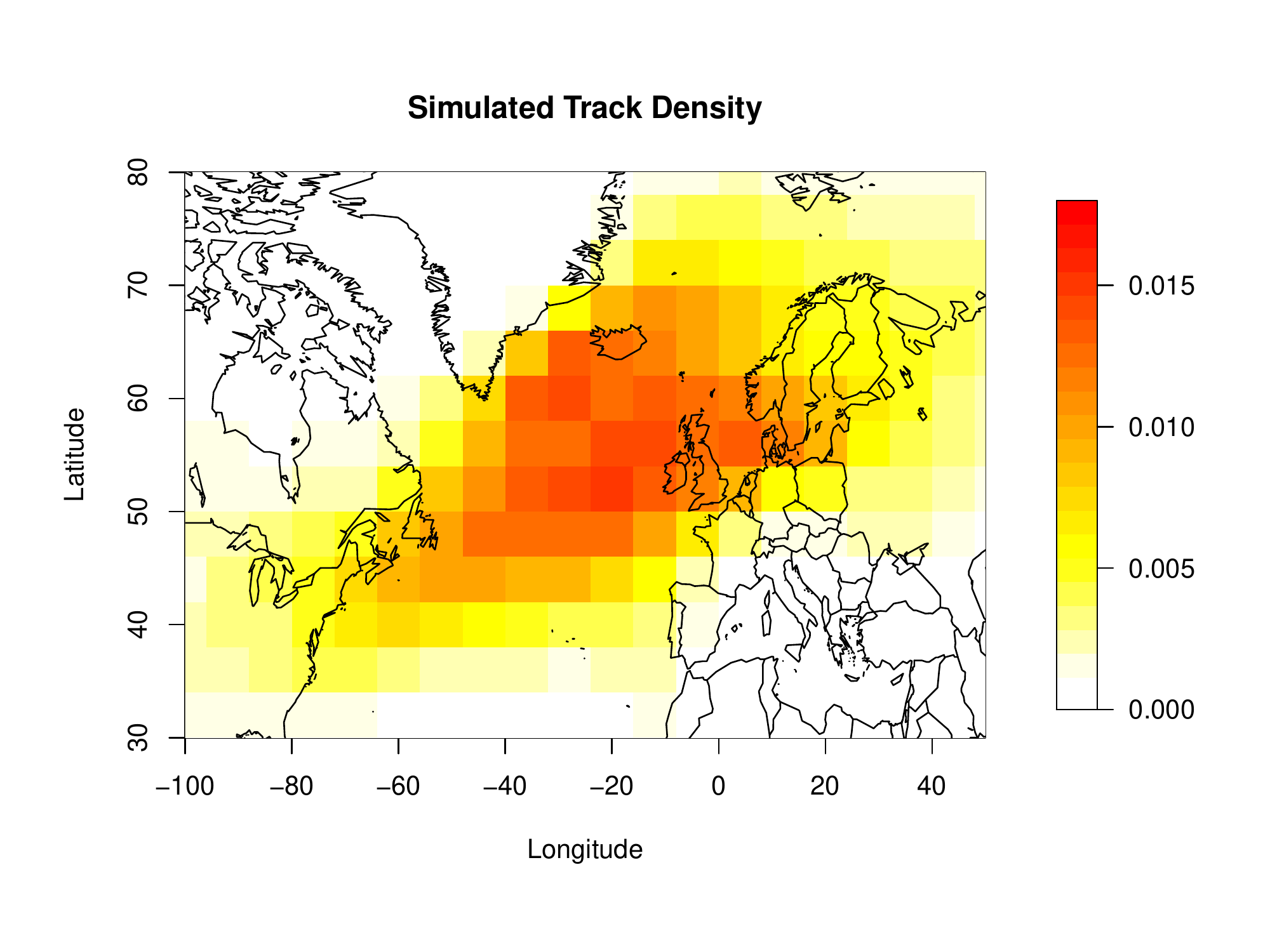}
\caption{Spatial density of a set of synthetic storm tracks simulated from the model.}
\label{fig:sim_spat_dens}
\end{figure}

We simulate $N=84,000$ synthetic storms from our model, which represents approximately $1,000$ years worth of storms, assuming the same average number of storms per year as observed in the data. Figure~\ref{fig:sim_spat_dens} shows the spatial density of the synthetic storms; it is clear that the model is capturing the spatial extent of the observed tracks as shown in Figure~\ref{fig:spat_dens}. We assess the model fit by constructing $95\%$ confidence intervals of the spatial density in each cell using a nonparametric bootstrap. Model estimates of the density were within these intervals more than $99\%$ of the time, which we deem sufficient evidence to suggest that the model is performing well.   \\
 
\begin{figure}[h!]
\centering
\includegraphics[width=12cm]{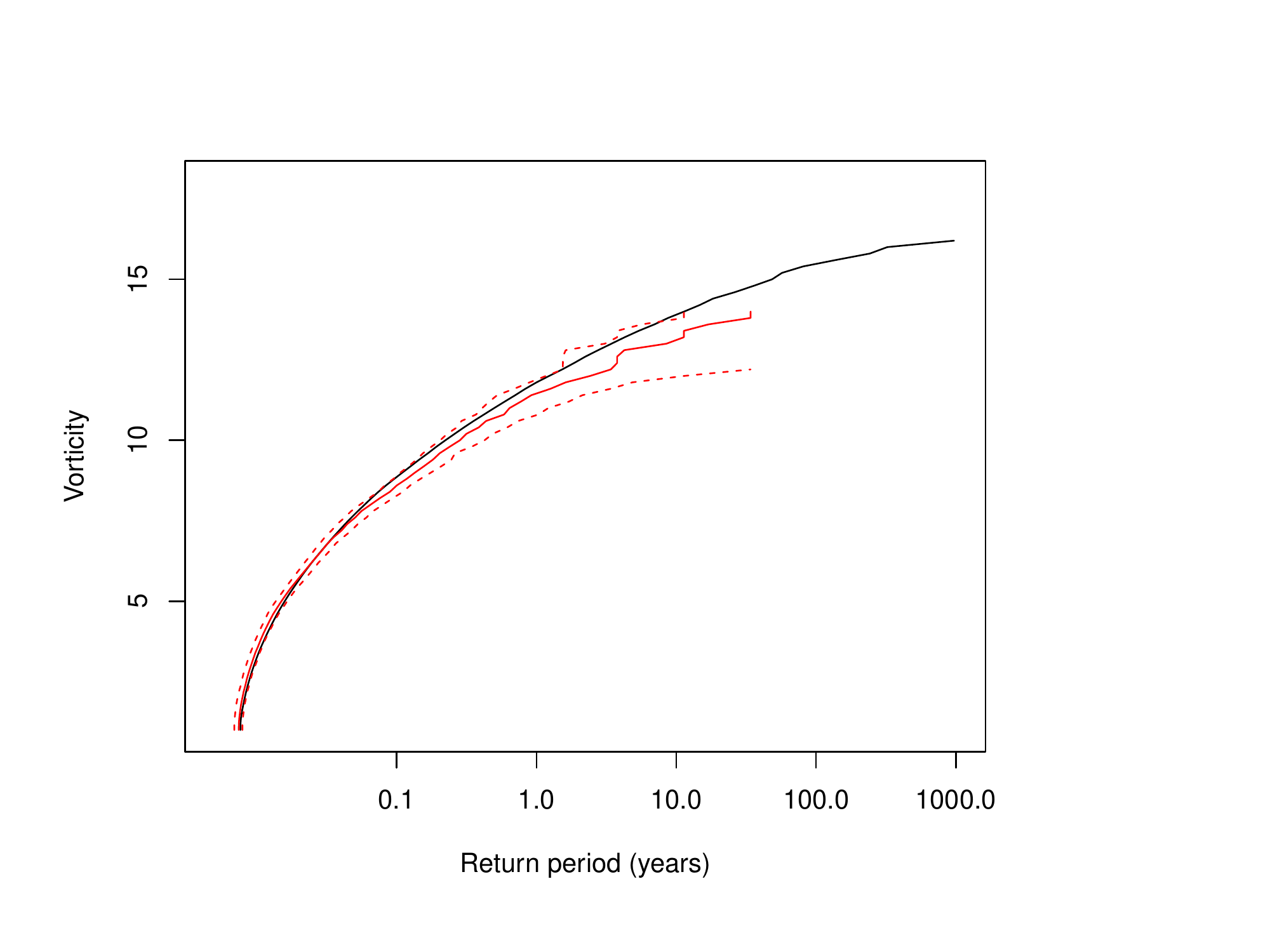}
\caption{The $r$-year return level curve for $\Gamma$ estimated empirically for a range of $r$ from the observed data (black) and from the Monte Carlo simulation from the model (red). The red dashed lines represent $95\%$ confidence intervals derived using a nonparametric bootstrap.}
\label{fig:retlev}
\end{figure}

Using the set of synthetic storms, we can estimate $\omega_r$, the $r$-year return level, that is, the vorticity value we expect to exceed once every $r$ years in $\Gamma$ such that $\Pr(\Omega_t > \omega_r | \boldsymbol{X}_t \in \Gamma) = 1/r$. Figure~\ref{fig:retlev} shows estimates of the $r$-year return level for $\Gamma$ from the observed data and using the Monte Carlo simulations from the model, where the model estimate falls within the $95\%$ confidence interval corresponding to the empirical estimate in the range of the data, obtained using a nonparametric bootstrap. However, our approach allows us to estimate return levels corresponding to events beyond the range of the data, meaning we can estimate the vorticity corresponding to a $100$-year or $1000$-year event, for example. \\
%
%
%

\begin{figure}
\centering
\includegraphics[width=12cm]{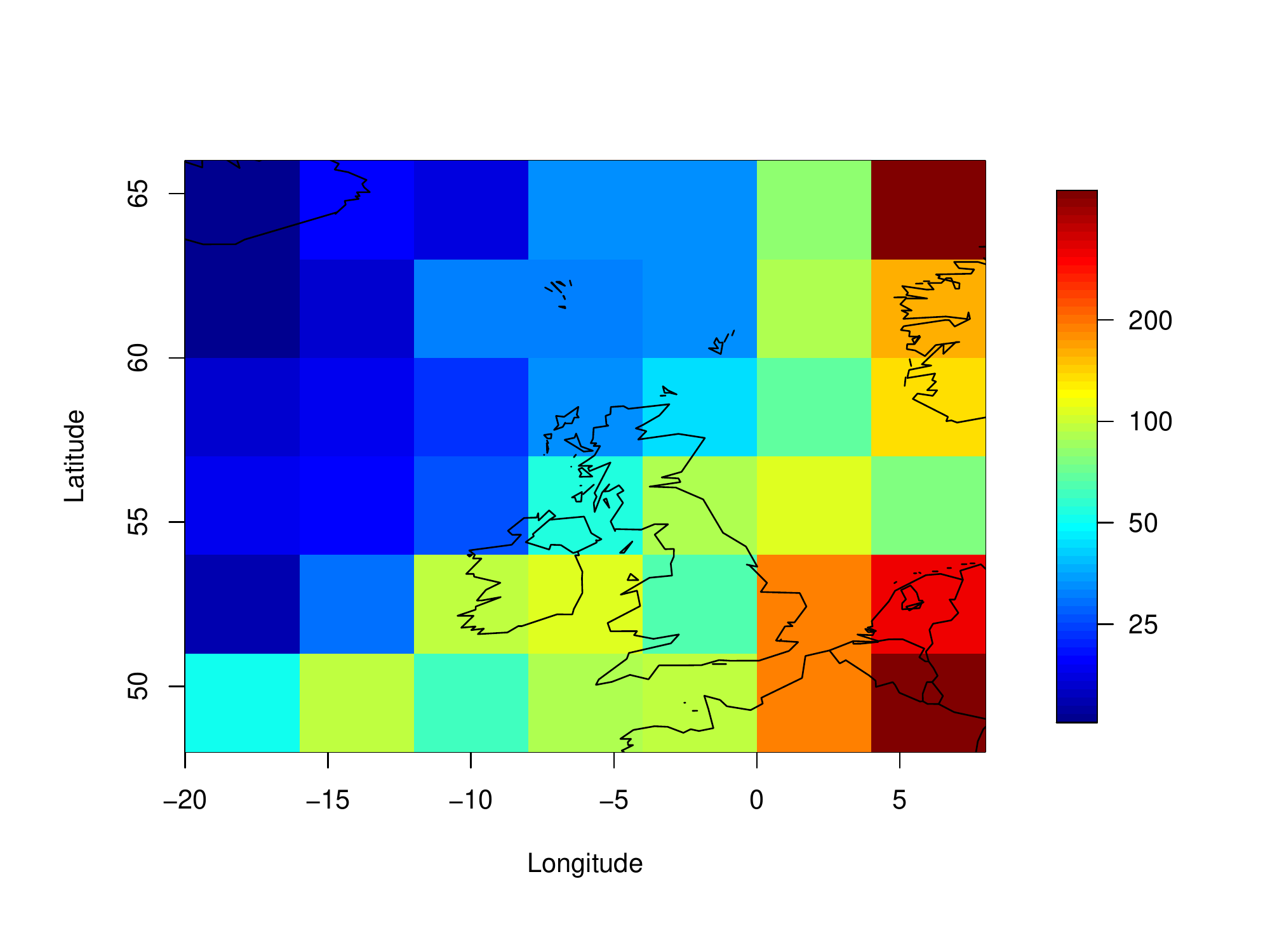}
\caption{Return period corresponding to a vorticity value of $\omega = 13.36 \times 10^{-5} s ^{-1}$ over space, estimated from the model through Monte Carlo simulation.}
\label{fig:herta_risk}
\end{figure}

\begin{figure}
\centering
\includegraphics[width=14cm]{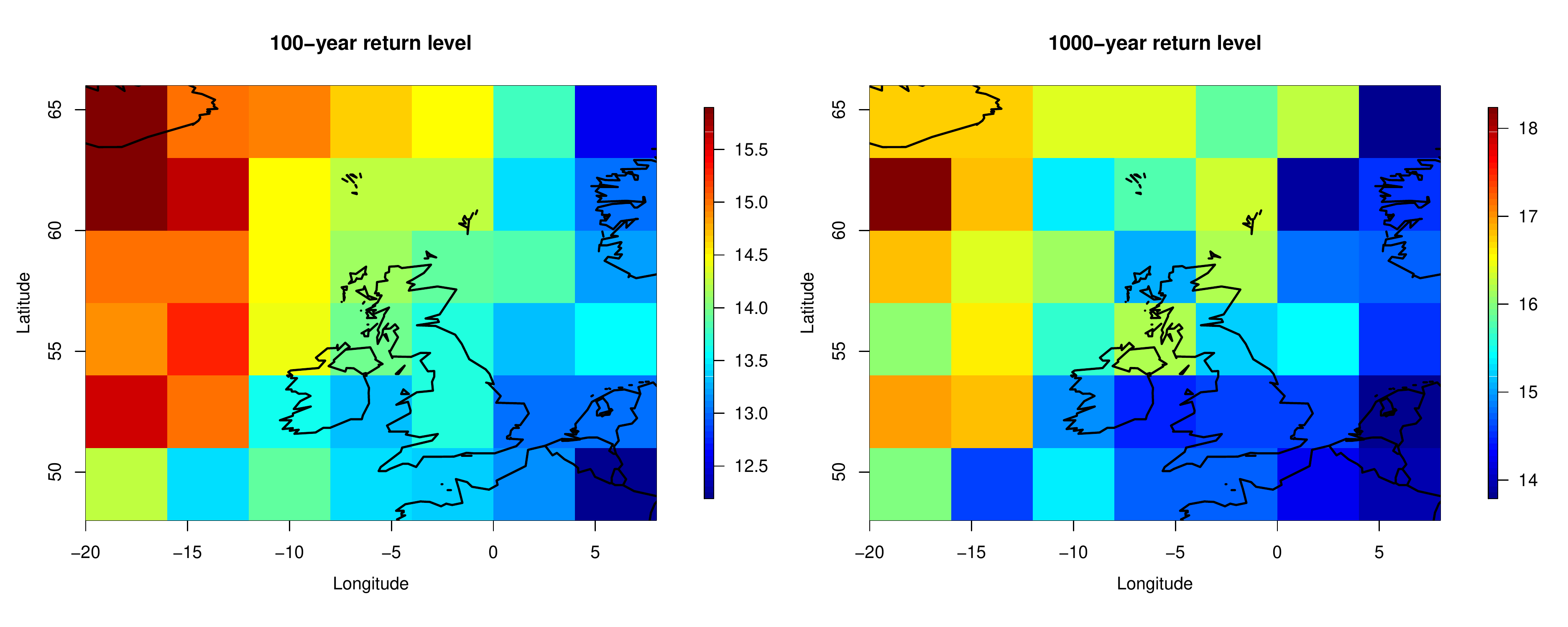}
\caption{The $100$-year (left) and $1,000$-year (right) return level over space, estimated from the model using Monte Carlo simulation.}
\label{fig:return_levels}
\end{figure}

One of the most destructive events to impact the UK in the last 35 years was Storm Herta, which caused approximately $\$1.5$ billion worth of damage after hitting Northern Europe in February 1990 and had a maximum observed vorticity of $13.36 \times 10^{-5} s ^{-1}$ \citep{roberts2014xws}. Through Monte Carlo simulation, we can estimate the return period of this event and assess the relative risks of extreme vorticity events over space. Figure~\ref{fig:herta_risk} shows the return period corresponding to an observation of $\omega=13.36 \times 10^{-5} s ^{-1}$ over different grid cells in a region containing the UK on a $4^{\circ} \times 3^{\circ}$ grid. Storm Herta reached its maximum vorticity at $\boldsymbol{x}=(1.89^\circ,55.63^\circ)$. The return period of this event in the cell containing this location (under this particular discretisation of space) is approximately $107$ years, whereas this event is much less rare in the north Atlantic, where return periods are $10-25$ years. This illustrates the strength of our model in assessing the relative risk of such extreme vorticity events over the spatial domain. Similarly, Figure~\ref{fig:return_levels} shows the $100$-year and $1,000$-year return levels estimated from the model under the same discretisation of space, again highlighting the increased probability of observing extreme vorticities in the north Atlantic compared to the UK and mainland Europe. 

\section{Discussion}
\label{sec:discussion}
We have developed a novel approach for simulating extra tropical cyclone tracks for the winter half year in the North Atlantic and European domain based on a stochastic model that captures the evolution and structure of observed storm systems. The storm track model is constructed by exploiting the spatio-temporal structure within observed storm tracks to initialise, propagate and terminate an individual storm, producing a synthetic track which reflects the key physical characteristics of these weather systems. Track climatologies derived from very large numbers of simulated tracks generated from the model reproduce well the observed spatial variation of the vorticity, lifecycle and tracks of storms. \\

For practitioners, our model is useful for improving risk assessment related to extreme weather driven by extratropical cyclone activity. The limited observed record means that risk assessments based on empirical evidence are highly uncertain and restricted to observed intensities, with no extrapolation possible beyond the range of the data. By supplementing the observed data with synthetic tracks from our physically-motivated model, probabilities of rare events can be calculated with increased confidence, including events of severity not yet seen,  which can assist in the design of defensive infrastructures. The robust validation of the simulated storm track climatology supports such an approach.  We find that the return period of storms with the same vorticity as storm Herta in February 1990, which caused approximately $\$1.5$ billion worth of damage in Northern Europe, is relatively frequent at $10-25$ years over the northern Atlantic, reducing to $\sim80$ years for Scotland and to $\sim200$ years for central France.\\

The effect of a changing climate on the climatology of the North Atlantic storm track and the potential future risk from extreme storms is of pressing concern but one which is also challenged by  the sampling issues addressed in this paper.  Earlier studies have indicated a poleward shift in the storm track with decreasing frequency and increasing intensity \citep{mccabe2001trends,bengtsson2006storm} whilst more recent studies have indicated that the future response is regionally and seasonally dependent \citep{zappa13futurextc} with uncertainty arising from competing physical processes and large internal variability in the climate system \citep{Shaw2016stormtrack}.  Our storm track model provides another tool to assess such changes and future risk through its application to storm track data from future climate simulations.\\

Improvements to the storm track model, both in terms of its physical realism and utility, include capturing the annual cycle of storm track characteristics. During the summer the preferred path of extratropical cyclones migrates northwards and returns southwards for winter, a feature the track model does not currently capture.  Furthermore, large scale modes of atmospheric variability, such as the North Atlantic Oscillation (NAO) are known to influence the path and frequency of extratropical storm tracks and the subsequent risk of extreme rainfall \citep{brown2017ukxppt}. The NAO, an anomalous dipole pressure pattern between the Icelandic low and the Azores high, significantly modifies the strength of the large scale westerly flow and location of the storm tracks. A positive NAO is associated with increased cyclonic activity in Northern Europe, while Southern Europe is typically susceptible to more storm events during a negative NAO phase \citep{mailier2006serial}. Exploring the annual cycle of track behaviour and their dependence an NAO represent interesting avenues for future work, enabling simulation of synthetic tracks specific to season and NAO phase. \\

\appendix
\section*{Acknowledgements}
The authors gratefully acknowledge the support of the EPSRC funded EP/H023151/1
STOR-i Centre for Doctoral Training, the Met Office and EDF Energy. We extend our thanks to Hugo Winter for helpful discussions and support. We thank Kevin Hodges for the storm track data. Simon Brown was supported by and Paul Sharkey partially supported by the Joint UK BEIS/Defra Met Office Hadley Centre Climate Programme (GA01101).

\section*{Appendix}
\section{Conditional kernel density estimation}
\label{App:sim_kernel}
Consider an arbitrary $d$-dimensional random vector $\boldsymbol{Z} = (Z_1, Z_2, \hdots, Z_d)$, which is observed $n$ times $\boldsymbol{z}^{(1)}, \boldsymbol{z}^{(2)}, \hdots, \boldsymbol{z}^{(n)}$. As a way of estimating $f(\boldsymbol{z})$, the joint probability density of $\boldsymbol{Z}$, we define the multivariate kernel density estimator as 
\begin{equation}
 \hat{f}(\boldsymbol{z}) = \frac{1}{n} \sum_{i=1}^{n} K_{\boldsymbol{H}} \left(\boldsymbol{z} - \boldsymbol{z}^{(i)} \right),
 \label{eq:mkde} 
\end{equation}
where $K$ is the kernel function and $\boldsymbol{H}$ denotes the bandwidth matrix which is symmetric and positive-definite. For our purposes, we choose $K$ to be the multivariate Gaussian density function
\begin{equation}
 K_{\boldsymbol{H}} (\boldsymbol{z}) = {(2 \pi)}^{-d/2} {|\boldsymbol{H}|}^{-1/2} \exp\left\{-\frac{1}{2} \boldsymbol{z}^{T} \boldsymbol{H}^{-1} \boldsymbol{z} \right\}
 \label{eq:kernel}
 \end{equation}
and the bandwidth matrix $\boldsymbol{H}$ chosen to be proportional to the rule-of-thumb selection of \citet{scott2015multivariate}. The bandwidth matrix $\boldsymbol{H}$ can be chosen to be diagonal or oriented. To simulate from the kernel density, we first sample uniformly a tuple $\boldsymbol{z}^{(i)}$, where $i \in \{1, \hdots, n\}$. We then simulate a vector $\tilde{\boldsymbol{z}}$, say, such that $\tilde{\boldsymbol{z}} \sim \text{MVN}(\boldsymbol{z}^{(i)},\boldsymbol{H})$. \\

Let $\boldsymbol{Z}$ be decomposed such that $\boldsymbol{Z} = (\boldsymbol{Z}_{-m},\boldsymbol{Z}_m)$. Consider the case when values $\boldsymbol{Z}_{-m} = \boldsymbol{z}_{-m}$ have been observed and we wish to estimate the conditional density of $\boldsymbol{Z}_{m}$ given these values. We can then define the conditional kernel density estimator as
\begin{equation} \hat{f}(\boldsymbol{z}_m | \boldsymbol{z}_{-m} ) = \sum_{i=1}^n w_i (\boldsymbol{z}_{-m}) K_{\boldsymbol{H}} \left(\boldsymbol{z}_m - \boldsymbol{z}_m^{(i)} \middle| \boldsymbol{z}_{-m} - \boldsymbol{z}_{-m}^{(i)} \right),
\label{eq:ckde}
\end{equation}
where 
\[
w_i (\boldsymbol{z}_{-m}) = \frac{K_{\boldsymbol{H}} \left( \boldsymbol{z}_{-m} - \boldsymbol{z}_{-m}^{(i)} \right)}{ \sum_{j=1}^{n} K_{\boldsymbol{H}} \left( \boldsymbol{z}_{-m} - \boldsymbol{z}_{-m}^{(j)} \right)},
\]
where $K_{\boldsymbol{H}} ( \cdot )$ is the multivariate Gaussian kernel function and $K_{\boldsymbol{H}} ( \cdot \mid \cdot )$ is the conditional Gaussian kernel function with bandwidth matrix $\boldsymbol{H}$ as defined in equation~(\ref{eq:kernel}). Let $\boldsymbol{H}$ be partitioned such that 
\[ \boldsymbol{H} = \begin{bmatrix}
\boldsymbol{H}_{m,m} & \boldsymbol{H}_{m,-m} \\ \boldsymbol{H}_{-m,m} & \boldsymbol{H}_{-m,-m}
\end{bmatrix}. \]

Conditional on having observed $\boldsymbol{z}_{-m}$, we choose a tuple $\boldsymbol{z}^{(i)}$ with probability $w_i (\boldsymbol{z}_{-m})$. Then we simulate 
\begin{equation}
 \boldsymbol{Z}_m | (\boldsymbol{Z}_{-m} = \boldsymbol{z}_{-m}) \sim \mathcal{N}(\bar{\boldsymbol{\mu}}, \bar{\boldsymbol{\Sigma}}),
 \label{eq:kersim}
\end{equation}
where $\bar{\boldsymbol{\mu}} = z_m^{(i)} + \boldsymbol{H}_{m,-m} \boldsymbol{H}_{-m,-m}^{-1} (\boldsymbol{z}_{-m} - \boldsymbol{z}_{-m}^{(i)}$) and $\bar{\boldsymbol{\Sigma}} = \boldsymbol{H}_{m,m} - \boldsymbol{H}_{m,-m}\boldsymbol{H}_{-m,-m}^{-1} \boldsymbol{H}_{-m,m}$. \\

\bibliographystyle{apalike}  
\nocite{*}
\bibliography{storm_writeup_arxiv}

\begin{thebibliography}{}

\bibitem[Akhtar et~al., 2014]{akhtar2014medicanes}
Akhtar, N., Brauch, J., Dobler, A., B{\'e}ranger, K., and Ahrens, B. (2014).
\newblock Medicanes in an ocean--atmosphere coupled regional climate model.
\newblock {\em Natural Hazards and Earth System Sciences}, 14:2189--2201.

\bibitem[Bengtsson et~al., 2006]{bengtsson2006storm}
Bengtsson, L., Hodges, K.~I., and Roeckner, E. (2006).
\newblock Storm tracks and climate change.
\newblock {\em Journal of Climate}, 19(15):3518--3543.

\bibitem[Bortot and Tawn, 1998]{bortot1998models}
Bortot, P. and Tawn, J.~A. (1998).
\newblock Models for the extremes of {M}arkov chains.
\newblock {\em Biometrika}, 85(4):851--867.

\bibitem[Brown, 2017]{brown2017ukxppt}
Brown, S.~J. (2017).
\newblock The drivers of variability in {UK} extreme rainfall.
\newblock {\em International Journal of Climatology}.

\bibitem[Casson and Coles, 2000]{casson2000simulation}
Casson, E. and Coles, S.~G. (2000).
\newblock Simulation and extremal analysis of hurricane events.
\newblock {\em Journal of the Royal Statistical Society: Series C (Applied
  Statistics)}, 49:227--245.

\bibitem[Coles, 2001]{coles2001introduction}
Coles, S.~G. (2001).
\newblock {\em An Introduction to Statistical Modeling of Extreme Values}.
\newblock Springer.

\bibitem[Coles et~al., 1999]{coles1999dependence}
Coles, S.~G., Heffernan, J.~E., and Tawn, J.~A. (1999).
\newblock Dependence measures for extreme value analyses.
\newblock {\em Extremes}, 2(4):339--365.

\bibitem[Davis and Mikosch, 2009]{davis2009extremogram}
Davis, R.~A. and Mikosch, T. (2009).
\newblock The extremogram: A correlogram for extreme events.
\newblock {\em Bernoulli}, 15(4):977--1009.

\bibitem[Davison and Smith, 1990]{davison1990models}
Davison, A.~C. and Smith, R.~L. (1990).
\newblock Models for exceedances over high thresholds (with discussion).
\newblock {\em Journal of the Royal Statistical Society. Series B
  (Methodological)}, 52(3):393--442.

\bibitem[Dee et~al., 2011]{dee2011era}
Dee, D.~P., Uppala, S.~M., Simmons, A.~J., Berrisford, P., Poli, P., Kobayashi,
  S., Andrae, U., Balmaseda, M.~A., Balsamo, G., and Bauer, P. (2011).
\newblock The {ERA}-{I}nterim reanalysis: Configuration and performance of the
  data assimilation system.
\newblock {\em Quarterly Journal of the Royal Meteorological Society},
  137(656):553--597.

\bibitem[Eastoe and Tawn, 2009]{eastoe2009modelling}
Eastoe, E.~F. and Tawn, J.~A. (2009).
\newblock Modelling non-stationary extremes with application to surface level
  ozone.
\newblock {\em Journal of the Royal Statistical Society: Series C (Applied
  Statistics)}, 58(1):25--45.

\bibitem[Economou et~al., 2014]{economou2014spatio}
Economou, T., Stephenson, D.~B., and Ferro, C. A.~T. (2014).
\newblock Spatio-temporal modelling of extreme storms.
\newblock {\em The Annals of Applied Statistics}, 8(4):2223--2246.

\bibitem[Hall and Jewson, 2007]{hall2007statistical}
Hall, T.~M. and Jewson, S. (2007).
\newblock Statistical modelling of {N}orth {A}tlantic tropical cyclone tracks.
\newblock {\em Tellus A}, 59(4):486--498.

\bibitem[Heffernan and Resnick, 2007]{heffernan2007limit}
Heffernan, J.~E. and Resnick, S.~I. (2007).
\newblock Limit laws for random vectors with an extreme component.
\newblock {\em The Annals of Applied Probability}, 17(2):537--571.

\bibitem[Heffernan and Tawn, 2004]{heffernan2004conditional}
Heffernan, J.~E. and Tawn, J.~A. (2004).
\newblock A conditional approach for multivariate extreme values (with
  discussion).
\newblock {\em Journal of the Royal Statistical Society: Series B (Statistical
  Methodology)}, 66(3):497--546.

\bibitem[Hodges, 1995]{hodges1995feature}
Hodges, K.~I. (1995).
\newblock Feature tracking on the unit sphere.
\newblock {\em Monthly Weather Review}, 123(12):3458--3465.

\bibitem[Hoskins and Hodges, 2002]{hoskins2002new}
Hoskins, B.~J. and Hodges, K.~I. (2002).
\newblock New perspectives on the {N}orthern {H}emisphere winter storm tracks.
\newblock {\em Journal of the Atmospheric Sciences}, 59(6):1041--1061.

\bibitem[Keef et~al., 2013]{keef2013estimation}
Keef, C., Papastathopoulos, I., and Tawn, J.~A. (2013).
\newblock Estimation of the conditional distribution of a multivariate variable
  given that one of its components is large: Additional constraints for the
  {H}effernan and {T}awn model.
\newblock {\em Journal of Multivariate Analysis}, 115:396--404.

\bibitem[Ledford and Tawn, 1996]{ledford1996statistics}
Ledford, A.~W. and Tawn, J.~A. (1996).
\newblock Statistics for near independence in multivariate extreme values.
\newblock {\em Biometrika}, 83(1):169--187.

\bibitem[Ledford and Tawn, 2003]{ledford2003diagnostics}
Ledford, A.~W. and Tawn, J.~A. (2003).
\newblock Diagnostics for dependence within time series extremes.
\newblock {\em Journal of the Royal Statistical Society: Series B (Statistical
  Methodology)}, 65(2):521--543.

\bibitem[Mailier et~al., 2006]{mailier2006serial}
Mailier, P.~J., Stephenson, D.~B., Ferro, C. A.~T., and Hodges, K.~I. (2006).
\newblock Serial clustering of extratropical cyclones.
\newblock {\em Monthly weather review}, 134(8):2224--2240.

\bibitem[McCabe et~al., 2001]{mccabe2001trends}
McCabe, G.~J., Clark, M.~P., and Serreze, M.~C. (2001).
\newblock Trends in northern hemisphere surface cyclone frequency and
  intensity.
\newblock {\em Journal of Climate}, 14(12):2763--2768.

\bibitem[Murray and Simmonds, 1991]{murray1991numerical}
Murray, R.~J. and Simmonds, I. (1991).
\newblock A numerical scheme for tracking cyclone centres from digital data.
\newblock {\em Australian Meteorological Magazine}, 39(3).

\bibitem[Papastathopoulos et~al., 2017]{papastathopoulos2017extreme}
Papastathopoulos, I., Strokorb, K., Tawn, J.~A., and Butler, A. (2017).
\newblock Extreme events of {M}arkov chains.
\newblock {\em Advances in Applied Probability}, 49(1):134--161.

\bibitem[Pickands, 1975]{pickands1975statistical}
Pickands, J. (1975).
\newblock Statistical inference using extreme order statistics.
\newblock {\em The Annals of Statistics}, 3(1):119--131.

\bibitem[Roberts et~al., 2014]{roberts2014xws}
Roberts, J.~F., Champion, A.~J., Dawkins, L.~C., Hodges, K.~I., Shaffrey,
  L.~C., Stephenson, D.~B., Stringer, M.~A., Thornton, H.~E., and Youngman,
  B.~D. (2014).
\newblock The {XWS} open access catalogue of extreme {E}uropean windstorms from
  1979 to 2012.
\newblock {\em Natural Hazards and Earth System Sciences}, 14:2487--2501.

\bibitem[Rumpf et~al., 2007]{rumpf2007stochastic}
Rumpf, J., Weindl, H., H{\"o}ppe, P., Rauch, E., and Schmidt, V. (2007).
\newblock Stochastic modelling of tropical cyclone tracks.
\newblock {\em Mathematical Methods of Operations Research}, 66(3):475--490.

\bibitem[Scott, 1992]{scott2015multivariate}
Scott, D.~W. (1992).
\newblock {\em Multivariate Density Estimation: Theory, Practice, and
  Visualization}.
\newblock John Wiley \& Sons.

\bibitem[Shapiro and Keyser, 1990]{shapiro1990fronts}
Shapiro, M.~A. and Keyser, D.~A. (1990).
\newblock {\em Fronts, jet streams, and the tropopause}.
\newblock US Department of Commerce, National Oceanic and Atmospheric
  Administration, Environmental Research Laboratories, Wave Propagation
  Laboratory.

\bibitem[Shaw et~al., 2016]{Shaw2016stormtrack}
Shaw, T.~A., Baldwin, M., Barnes, E.~A., Caballero, R., Garfinkel, C.~I.,
  Hwang, Y.-T., Li, C., O'Gorman, P.~A., Riviere, G., Simpson, I.~R., and
  Voigt, A. (2016).
\newblock Storm track processes and the opposing influences of climate change.
\newblock {\em Nature Geoscience}, 9:656--664.

\bibitem[Sienz et~al., 2010]{sienz2010extreme}
Sienz, F., Schneidereit, A., Blender, R., Fraedrich, K., and Lunkeit, F.
  (2010).
\newblock Extreme value statistics for {N}orth {A}tlantic cyclones.
\newblock {\em Tellus A}, 62(4):347--360.

\bibitem[Silverman, 1986]{silverman1986density}
Silverman, B.~W. (1986).
\newblock {\em Density estimation for statistics and data analysis}.
\newblock CRC Press.

\bibitem[Ulbrich et~al., 2009]{ulbrich2009extra}
Ulbrich, U., Leckebusch, G.~C., and Pinto, J.~G. (2009).
\newblock Extra-tropical cyclones in the present and future climate: a review.
\newblock {\em Theoretical and Applied Climatology}, 96(1-2):117--131.

\bibitem[Winter and Tawn, 2017]{winter2016kth}
Winter, H.~C. and Tawn, J.~A. (2017).
\newblock kth-order {M}arkov extremal models for assessing heatwave risks.
\newblock {\em Extremes}, 20(2):393--415.

\bibitem[Wood, 2006]{wood2006generalized}
Wood, S. (2006).
\newblock {\em Generalized {a}dditive {m}odels: an {i}ntroduction with R}.
\newblock CRC press.

\bibitem[Zappa et~al., 2013]{zappa13futurextc}
Zappa, G., Shaffrey, L.~C., Hodges, K.~I., Sansom, P.~G., and Stephenson, D.~B.
  (2013).
\newblock A multi-model assessment of future projections of {N}orth {A}tlantic
  and {E}uropean extratropical cyclones in the {CMIP5} climate models.
\newblock {\em Journal of Climate}, 26(16):5846--5862.

\end{thebibliography}
\end{document}